\documentclass[sigconf]{acmart}

\usepackage[ruled,vlined,linesnumbered]{algorithm2e}
\usepackage{balance}
\usepackage{multirow}
\usepackage{amsfonts}
\usepackage{colortbl}
\usepackage{framed}
\usepackage{url}
\usepackage{xspace}
\usepackage{listings}
\usepackage{enumitem}
\usepackage{overpic}
\usepackage{subcaption}
\usepackage{graphicx}
\usepackage{grffile}
\usepackage{makecell}

\renewcommand\footnotetextcopyrightpermission[1]{}

\copyrightyear{2025}
\acmYear{2025}
\setcopyright{acmlicensed}
\acmConference[SIGIR '25]{Proceedings of the 48th International ACM SIGIR Conference on Research and Development in Information Retrieval}{July 13--18, 2025}{Padua, Italy}
\acmBooktitle{Proceedings of the 48th International ACM SIGIR Conference on Research and Development in Information Retrieval (SIGIR '25), July 13--18, 2025, Padua, Italy}
\acmDOI{10.1145/3726302.3730277}
\acmISBN{979-8-4007-1592-1/2025/07}

\begin{document}

\title{Benchmarking Recommendation, Classification, and Tracing Based on Hugging Face Knowledge Graph}

\author{Qiaosheng Chen}
\orcid{0009-0002-0610-7725}
\affiliation{
  \institution{Nanjing University}
  \department{State Key Laboratory for Novel Software Technology}
  \city{Nanjing}
  \state{Jiangsu}
  \country{China}
}
\email{qschen@smail.nju.edu.cn}

\author{Kaijia Huang}
\orcid{0009-0004-0359-7946}
\affiliation{
  \institution{Nanjing University}
  \department{State Key Laboratory for Novel Software Technology}
  \city{Nanjing}
  \state{Jiangsu}
  \country{China}
}
\email{atyou813@gmail.com}

\author{Xiao Zhou}
\orcid{0009-0008-1132-6408}
\affiliation{
  \institution{Nanjing University}
  \department{State Key Laboratory for Novel Software Technology}
  \city{Nanjing}
  \state{Jiangsu}
  \country{China}
}
\email{xzhou@smail.nju.edu.cn}

\author{Weiqing Luo}
\orcid{0009-0004-8041-3258}
\affiliation{
  \institution{Nanjing University}
  \department{State Key Laboratory for Novel Software Technology}
  \city{Nanjing}
  \state{Jiangsu}
  \country{China}
}
\email{wqluo@smail.nju.edu.cn}

\author{Yuanning Cui}
\orcid{0000-0002-9113-0155}
\affiliation{
  \institution{Nanjing University}
  \department{State Key Laboratory for Novel Software Technology}
  \city{Nanjing}
  \state{Jiangsu}
  \country{China}
}
\email{yncui.nju@gmail.com}

\author{Gong Cheng}
\authornote{Corresponding author}
\orcid{0000-0003-3539-7776}
\affiliation{
  \institution{Nanjing University}
  \department{State Key Laboratory for Novel Software Technology}
  \city{Nanjing}
  \state{Jiangsu}
  \country{China}
}
\email{gcheng@nju.edu.cn}

\renewcommand{\shortauthors}{Qiaosheng Chen et al.}

\begin{abstract}

The rapid growth of open source machine learning (ML) resources, such as models and datasets, has accelerated IR research. However, existing platforms like Hugging Face do not explicitly utilize structured representations, limiting advanced queries and analyses such as tracing model evolution and recommending relevant datasets. To fill the gap, we construct $\mathsf{HuggingKG}$, the first large-scale knowledge graph built from the Hugging Face community for ML resource management. With 2.6 million nodes and 6.2 million edges, $\mathsf{HuggingKG}$ captures domain-specific relations and rich textual attributes. It enables us to further present $\mathsf{HuggingBench}$, a multi-task benchmark with three novel test collections for IR tasks including resource recommendation, classification, and tracing. Our experiments reveal unique characteristics of $\mathsf{HuggingKG}$ and the derived tasks. Both resources are publicly available, expected to advance research in open source resource sharing and management.

\end{abstract}




\maketitle

\section{Introduction}
\label{sec:introduction}


The proliferation of open source models and datasets has empowered researchers and developers to build on existing AI tools, driving innovation across diverse domains.
It also highlights the critical role of resource sharing platforms in the advancement of AI research, which requires efficient frameworks to share, search, and manage software resources to ensure equitable access and foster collaboration~\cite{DatasetAudit, DatasetSearchSurvey, DatasetDiscoverySurvey, DEKR, tse23}. In this context, Hugging Face has emerged as a popular platform, democratizing access to cutting-edge ML resources and allowing users to find open models and datasets.

\textbf{Motivation.}
Existing platforms for open source resource sharing, such as GitHub and Hugging Face, rely primarily on keyword-based search and simplistic metadata tagging~\cite{datasets}. Although these platforms provide access to vast collections of models and datasets, they \emph{do not leverage semantic relations}~(e.g.,~model evolution, task dependencies, and user collaboration patterns) between resources during searches. As a result, they are \emph{unable to integrate structural information} to recommend and manage resources effectively. This unstructured paradigm severely limits support for advanced queries and analyses, such as tracing a model's evolution history, identifying relevant datasets for a specific task, or recommending underutilized resources based on structural dependencies.

\begin{figure*}[t]
    \centering
    \includegraphics[width=0.9\linewidth]{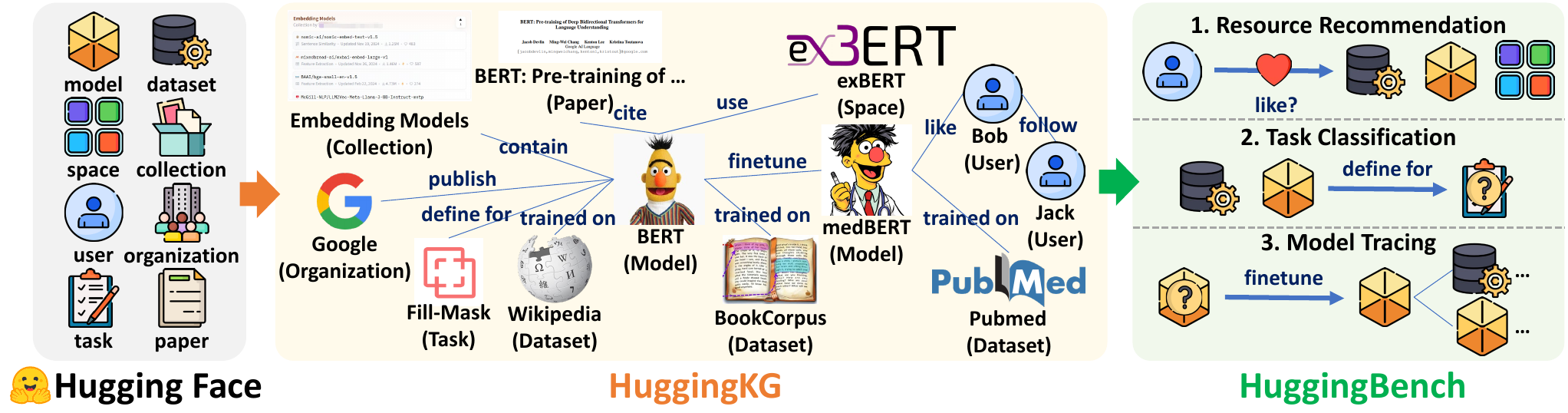}
    \caption{Illustration of $\mathsf{HuggingKG}$ and $\mathsf{HuggingBench}$.}
    \label{fig:huggingkg}
\end{figure*}

Knowledge graphs (KGs) offer a proven solution to this limitation. By explicitly representing entities as nodes and connecting them with edges representing typed relations, KGs enable sophisticated querying and analysis, as demonstrated in domains such as encyclopedias~\cite{Wikidata}, academia~\cite{OAG} and e-commerce~\cite{Amazon}. In the context of resource management, a structured KG can unify separated metadata, resource dependencies, and user interactions into a single heterogeneous graph. This representation supports complex tasks such as recommendation~\cite{DEKR, tse23, paper2repo}, link prediction~\cite{issue-PR-link-prediction}, and node classification~\cite{OAGBench}. Therefore, our work aims to \emph{transform resource sharing platforms into KGs that enhance discovery, reproducibility, and management of open source resources}.

\textbf{Our Resources.}  
We construct two foundational resources as illustrated in Figure~\ref{fig:huggingkg}: $\mathsf{HuggingKG}$, a large-scale ML resource KG built on the Hugging Face community, and $\mathsf{HuggingBench}$, a multi-task benchmark designed to evaluate practical challenges in open source resource management. With more than \emph{2.6 million nodes} and \emph{6.2 million triples}, $\mathsf{HuggingKG}$ represents the largest publicly available KG in this domain~(to our best knowledge). It integrates ML-related entities (e.g.,~\texttt{Model}, \texttt{Dataset}, \texttt{User}, \texttt{Task}) with ML-specific relations such as model evolution~(e.g.,~\texttt{Adapter}, \texttt{Finetune}) and user interactions~(e.g.,~\texttt{Like}, \texttt{Follow}), while incorporating rich textual attributes~(e.g.,~extended descriptions). $\mathsf{HuggingBench}$ includes \emph{the first cross-type resource recommendation test collection in the ML resource domain} and \emph{the first model tracing test collection}, novel domain-specific tasks that existing benchmarks cannot support. We analyze the unique properties of $\mathsf{HuggingKG}$ and perform extensive evaluations with $\mathsf{HuggingBench}$ to reveal their special characteristics, which differ from traditional KGs or tasks in domains such as encyclopedias, academia, or e-commerce.

The main contributions of this work are as follows.  
\begin{itemize}[leftmargin=*]  
    \item $\mathsf{HuggingKG}$: The first KG for ML resources on Hugging Face, featuring the largest scale among comparable works, and uniquely capturing domain-specific relations related to model evolution and user interaction as well as rich textual attributes.
    \item $\mathsf{HuggingBench}$: A new multi-task benchmark including novel test collections for cross-type resource recommendation, task classification, and model lineage tracing, addressing unmet needs in open source resource management.
\end{itemize}

\textbf{Availability.}
$\mathsf{HuggingKG}$ and $\mathsf{HuggingBench}$ are available on Hugging Face.\footnote{\url{https://huggingface.co/collections/cqsss/huggingbench-67b2ee02ca45b15e351009a2}} The code for constructing $\mathsf{HuggingKG}$ and reproducing the experiments is available on GitHub.\footnote{\url{https://github.com/nju-websoft/HuggingBench}} Both resources are licensed under Apache License 2.0.

\section{Related Work}
\label{sec:related-work}


\textbf{KGs for Resource Management.}
KGs have been extensively used to represent and analyze complex relationships in various domains, including open source resource management. Previous works such as DEKR~\cite{DEKR} and MLTaskKG~\cite{tse23} have constructed KGs to support recommendation tasks by capturing relationships among ML resources. DEKR~\cite{DEKR} primarily relies on description enhancement for ML method recommendation. MLTaskKG~\cite{tse23} constructs an AI task-model KG by integrating static data to support task-oriented ML/DL library recommendation. However, both approaches focus on static attributes and a narrow set of relations, failing to capture dynamic user interactions and inter-\texttt{Model} relations. In contrast, as shown in Table~\ref{tab:kg-bench-comparison}, our proposed $\mathsf{HuggingKG}$ is built on the rich metadata provided by Hugging Face, offering a large-scale KG with a more extensive set of relations. In addition to generic relations (e.g.,~\texttt{Defined For}, \texttt{Cite}), $\mathsf{HuggingKG}$ incorporates multiple inter-\texttt{Model} relations (i.e.,~\texttt{Adapter}, \texttt{Finetune}, \texttt{Merge}, and \texttt{Quantize}) and captures user interaction signals (i.e.,~\texttt{Publish}, \texttt{Like}, and \texttt{Follow}). \emph{This enriched structure facilitates a deeper analysis of ML resources and supports more effective recommendation strategies.}

\textbf{KG-based Benchmarks.}
Various benchmark datasets have been proposed to evaluate KG-based tasks. For example, OAG-Bench~\cite{OAGBench} provides a human-curated benchmark for academic graph mining, focusing on citation and collaboration networks. In the domain of open source resource management, our $\mathsf{HuggingBench}$ benchmark distinguishes itself by providing datasets for three IR tasks: resource recommendation, task classification, and model tracing.

For \textbf{resource recommendation}, paper2repo~\cite{paper2repo} introduces a distant-supervised recommender system that matches papers with related code repositories. However, it incorporates a limited range of entity types that are insufficient to build fine-grained interdependencies. Xu et al.~\cite{RepoRecommendation} leverages multi-modal features from developers’ sequential behaviors and repository text to generate relevant and tailored suggestions for developers, yet it does not explicitly construct or exploit a structured KG. In contrast, as shown in Table~\ref{tab:kg-bench-comparison}, \emph{$\mathsf{HuggingBench}$ benefits from the inherent structure of $\mathsf{HuggingKG}$ that captures rich relational data for recommendation}.

Furthermore, GRETA~\cite{GRETA} and recent efforts in automated categorization~\cite{EASE24, ESEM24} address specific \textbf{tagging/classification} tasks. GRETA~\cite{GRETA} constructs an Entity Tag Graph (ETG) using the cross-community knowledge from GitHub and Stack Overflow, and uses an iterative random walk with restart algorithm to automatically assign tags to repositories. \emph{$\mathsf{HuggingKG}$ integrates richer textual descriptions and metadata to construct a graph that encapsulates fine-grained relationships among models and datasets, thereby facilitating multi-label task classification for ML resources.}

\begin{table*}[t]
\centering
\caption{Comparison between KGs and benchmarks on open source resource management.}
\label{tab:kg-bench-comparison}
\resizebox{\textwidth}{!}{
\begin{tabular}{@{}lcrrrrcccc@{}}
\toprule
 & \textbf{Source} & \textbf{\#Nodes} & \textbf{\#Types} & \textbf{\#Relations} & \textbf{\#Edges} & \textbf{Key Entities \& (Attributes)} & \textbf{Model Evolution} & \textbf{User Interaction} & \textbf{Tasks} \\
\midrule
DEKR~\cite{DEKR} & \begin{tabular}{@{}c@{}}Open academic platforms\\(e.g.,~PapersWithCode, GitHub)\end{tabular} & 17,483 & 5 & 23 & 117,245 & \begin{tabular}{@{}c@{}}\texttt{Dataset}, \texttt{Method}\\(Description)\end{tabular} & No & No & Recommendation \\
\specialrule{0em}{1.5pt}{1.5pt}
MLTaskKG~\cite{tse23} & \begin{tabular}{@{}c@{}}PapersWithCode,\\ ML/DL Papers,\\ ML/DL Framework Docs\end{tabular} & 159,310 & 16 & 39 & 628,045 & \begin{tabular}{@{}c@{}}\texttt{Task}, \texttt{Model},\\ \texttt{Model Implementation}\end{tabular} & No & No & Recommendation \\
\specialrule{0em}{1.5pt}{1.5pt}
paper2repo~\cite{paper2repo} & \begin{tabular}{@{}c@{}}GitHub,\\ Microsoft Academic\end{tabular} & 39,600 & 2 & - & - & \texttt{Paper}, \texttt{Repository} & No & \begin{tabular}{@{}c@{}}Yes\\(\texttt{Star})\end{tabular} & Recommendation \\
\specialrule{0em}{1.5pt}{1.5pt}
GRETA~\cite{GRETA} & \begin{tabular}{@{}c@{}}GitHub,\\ Stack Overflow\end{tabular} & 707,891 & 4 & - & - & \texttt{Repository}, \texttt{Tag} & No & \begin{tabular}{@{}c@{}}Yes\\(\texttt{Search}, \texttt{Raise}, \texttt{Answer})\end{tabular} & Tag Assignment \\
\specialrule{0em}{1.5pt}{1.5pt}
AIPL(Facebook/React)~\cite{issue-PR-link-prediction} & GitHub & 97,556 & 4 & 9 & 196,834 & \begin{tabular}{@{}c@{}}\texttt{Issue}, \texttt{PR},\\ \texttt{Repository}, \texttt{User}\end{tabular} & No & Yes & Issue-PR Link Prediction \\
\specialrule{0em}{1.5pt}{1.5pt}
AIPL(vuejs/vue)~\cite{issue-PR-link-prediction} & GitHub & 49,200 & 4 & 9 & 95,160 & \begin{tabular}{@{}c@{}}\texttt{Issue}, \texttt{PR},\\ \texttt{Repository}, \texttt{User}\end{tabular} & No & Yes & Issue-PR Link Prediction \\
\midrule
$\mathsf{HuggingKG}$ \& $\mathsf{HuggingBench}$ & \textbf{Hugging Face} & \textbf{2,614,270} & \textbf{8} & \textbf{30} & \textbf{6,246,353} & \textbf{\begin{tabular}{@{}c@{}}\texttt{Model}, \texttt{Dataset},\\ \texttt{User}, \texttt{Task}\\(Description)\end{tabular}} & \textbf{\begin{tabular}{@{}c@{}}Yes\\(\texttt{Finetune}, \texttt{Adapter},\\ \texttt{Merge}, \texttt{Quantize})\end{tabular}} & \textbf{\begin{tabular}{@{}c@{}}Yes\\(\texttt{Publish}, \texttt{Like}, \texttt{Follow})\end{tabular}} & \textbf{\begin{tabular}{@{}c@{}}Recommendation,\\ Classification, Tracing\end{tabular}} \\
\bottomrule
\end{tabular}
}
\end{table*}

Recent work by Bai et al. ~\cite{issue-PR-link-prediction} uses a knowledge-aware heterogeneous graph learning approach to \textbf{predict links} between issues and pull requests on GitHub, effectively capturing complex relational information through metapath aggregation. However,
it remains confined to linking \texttt{Issue}–\texttt{PR} pairs and does not address the broader challenge of tracking model evolution across ML resources. \emph{The novel model tracing task in $\mathsf{HuggingBench}$ not only pioneers the exploration of inter-\texttt{Model} relations, but also provides practical insights into the evolution, reuse, and optimization of ML models}, thereby supporting more informed decision-making in real-world open source resource management.

\section{$\mathsf{HuggingKG}$ Knowledge Graph}
\label{sec:huggingkg}


\subsection{KG Construction}
\label{sec:kg-construction}
The construction of $\mathsf{HuggingKG}$ follows a principled process that includes defining nodes and edges, crawling and converting data from the Hugging Face community website, and performing data verification and cleaning.

\begin{figure}[t]
    \centering
    \includegraphics[width=\linewidth]{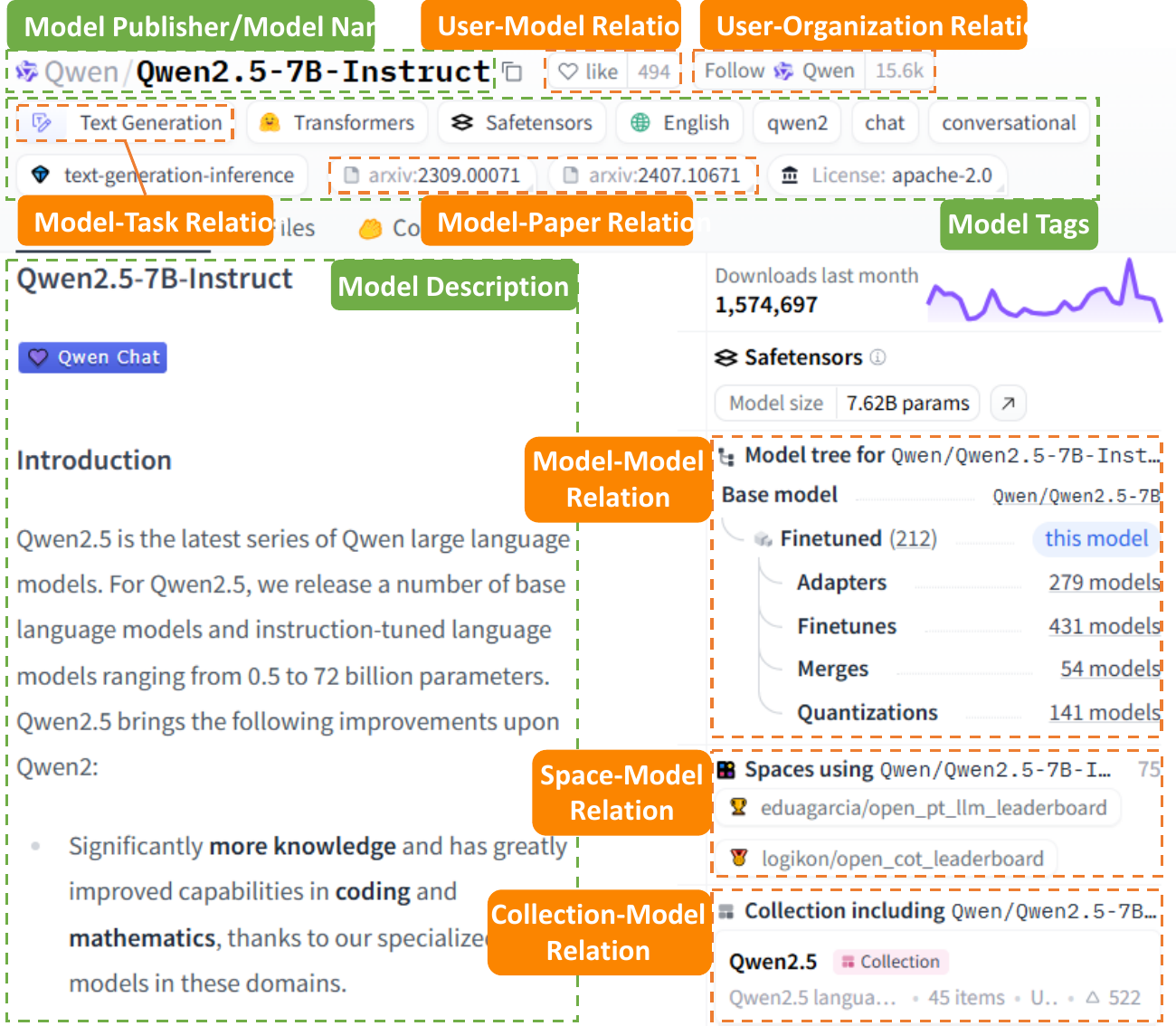}
    \caption{An example model page on Hugging Face.}
    \label{fig:example}
\end{figure}

\textbf{Schema Definition.}
The nodes and edges in $\mathsf{HuggingKG}$ are defined based on our meticulous analysis of the Hugging Face website and general IR needs in real-world scenarios. Figure~\ref{fig:example} shows an example model page of $\mathtt{Qwen/Qwen2.5\text{-}7B\text{-}Instruct}$\footnote{\url{https://huggingface.co/Qwen/Qwen2.5-7B-Instruct}} on the Hugging Face website. We can intuitively see that the key attributes of a \texttt{Model} include its name, publisher, tags, and text description on the model card, etc. The key relations that can be observed on the page include \texttt{Finetune} between \texttt{Model}s and \texttt{Like} between \texttt{User} and \texttt{Model}, etc.

\begin{figure*}
    \centering
    \includegraphics[width=0.8\linewidth]{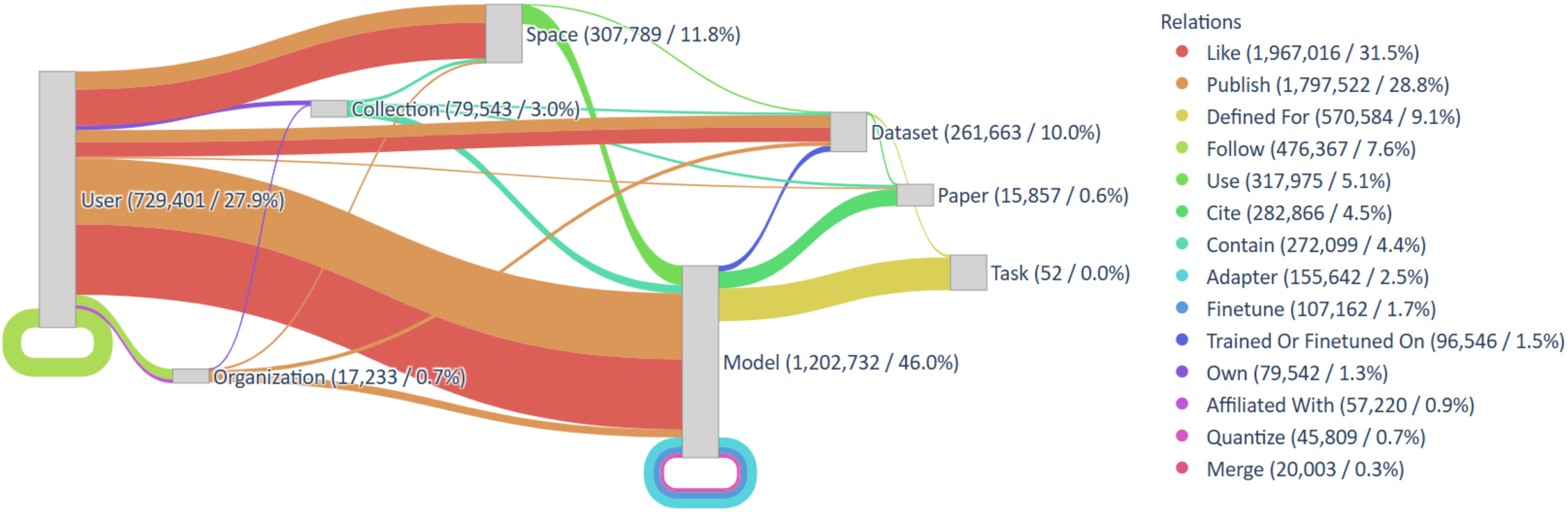}
    \caption{The schema graph of $\mathsf{HuggingKG}$, along with the quantity and proportion of each node and edge type.}
    \label{fig:schema}
\end{figure*}

Through an analysis of pages such as models, datasets, and spaces, we identify 8 types of nodes and 30 types of edges between them, as illustrated in Figure~\ref{fig:schema}. In addition, we determine the key attributes associated with each node type. For \texttt{Model}, \texttt{Dataset}, \texttt{Space}, and \texttt{Collection}, we adopt the ``publisher/name'' format used by Hugging Face serving as the primary identifier. To address potential name duplication across different node types (e.g.,~a model\footnote{\url{https://huggingface.co/aai540-group3/diabetes-readmission}} and a dataset\footnote{\url{https://huggingface.co/datasets/aai540-group3/diabetes-readmission}} sharing the same ``publisher/name''), we introduce a type prefix for each node. For example, a \texttt{Model} might be represented as $\mathtt{model:Qwen/Qwen2.5\text{-}7B\text{-}Instruct}$, ensuring its uniqueness. This string is used as a unique identifier to detect and prevent duplication during subsequent data crawling and processing.

\textbf{Data Crawling and Conversion.}  
The data crawling process is performed utilizing the \texttt{huggingface\_hub} library\footnote{\url{https://huggingface.co/docs/hub/index}} and issuing requests to the relevant API endpoints.\footnote{\url{https://huggingface.co/docs/hub/api}}

For \texttt{Model}, \texttt{Dataset}, and \texttt{Space}, we retrieve their complete lists, as well as their key attributes (e.g.,~tags, download count) and edges through functions in the \texttt{huggingface\_hub} library. Specifically, to capture the complete README file for each \texttt{Model} or \texttt{Dataset} node, we send a request to download and parse the \texttt{README.md} file, storing its content as the description field in text format. The list of \texttt{Collection} and additional metadata for these four node types are obtained by batch API requests.

The lists of \texttt{User} and \texttt{Organization} are extracted from the ``publisher'' fields of various nodes, including \texttt{Model}, \texttt{Dataset}, \texttt{Space}, \texttt{Collection}, and \texttt{Paper}. \texttt{User} and \texttt{Organization} nodes are distinguished by their profiles from the API endpoints, and edges such as \texttt{Follow} and \texttt{Affiliated With} are also captured.

\texttt{Paper} nodes are identified through \texttt{arxiv:} tags present in \texttt{Model}, \texttt{Dataset}, and \texttt{Collection} nodes. Detailed metadata, such as title, abstract, publication date, and authorship, is retrieved through API calls. For authors registered as \texttt{User}, we established edges between \texttt{User} nodes and their associated \texttt{Paper} nodes.

\texttt{Task} nodes are identified from three primary sources: ``tasks'' fields in \texttt{Dataset}, ``pipeline tags'' fields in \texttt{Model}, and direct definitions from the API. Edges between \texttt{Task} nodes and \texttt{Model} or \texttt{Dataset} nodes, are established by parsing metadata fields, allowing us to align models and datasets with the tasks they serve for.

\textbf{Data Verification and Cleaning.}  
After data crawling is completed, we verify and clean all nodes and edges to ensure the accuracy and completeness of $\mathsf{HuggingKG}$. The verification process involves scanning all collected edges and checking whether the nodes involved in each edge exist within the set of collected nodes. Some models may have been deleted by the publisher, but edges involving those models (e.g., a model finetuned from the original) may still appear on Web pages. If any invalid edges are detected, they are removed. The data cleaning process primarily focuses on eliminating outliers in node attributes and removing invalid characters from text fields.

\textbf{Time and Space Cost.}
To reduce time overhead, the construction process employs multi-threaded parallel processing at each step. Consequently, the entire process takes approximately 20 hours and the storage of node attributes and edges in the graph amounts to around 5.8 GB. This indicates that the graph can be updated daily. The current version of $\mathsf{HuggingKG}$ is constructed with data collected on December 15, 2024.

\begin{figure}
    \centering
    \begin{subfigure}{\linewidth}
        \includegraphics[width=\linewidth]{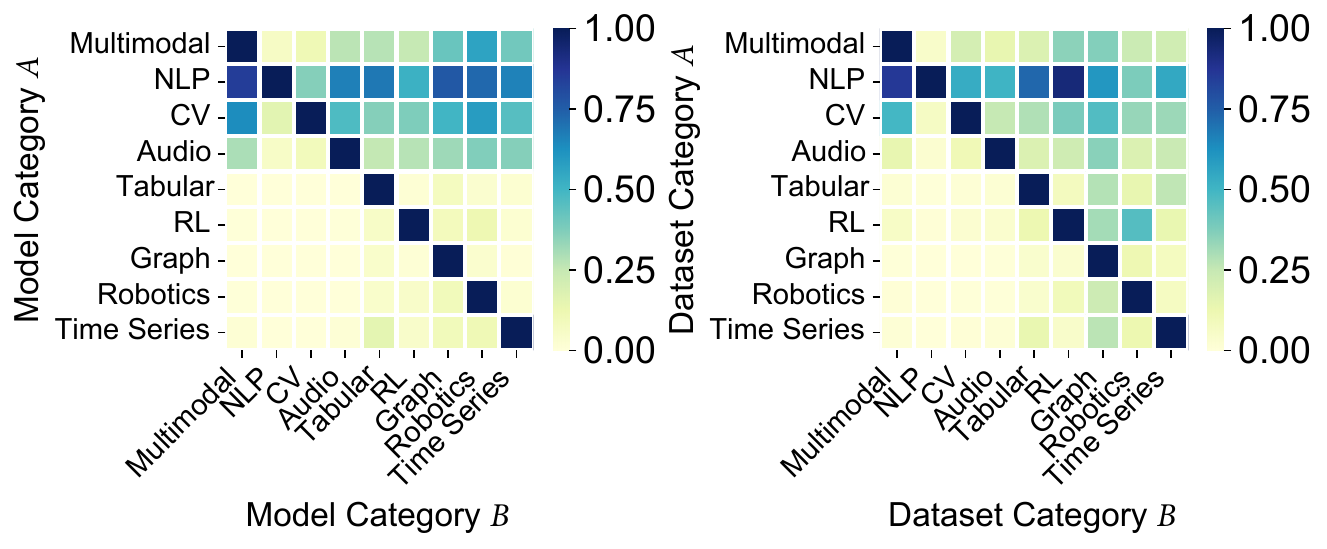}
        \caption{$P(A|B)$ for within-type nodes.}
        \label{fig:colike-heatmap1}
    \end{subfigure}
    \begin{subfigure}{\linewidth}
        \includegraphics[width=\linewidth]{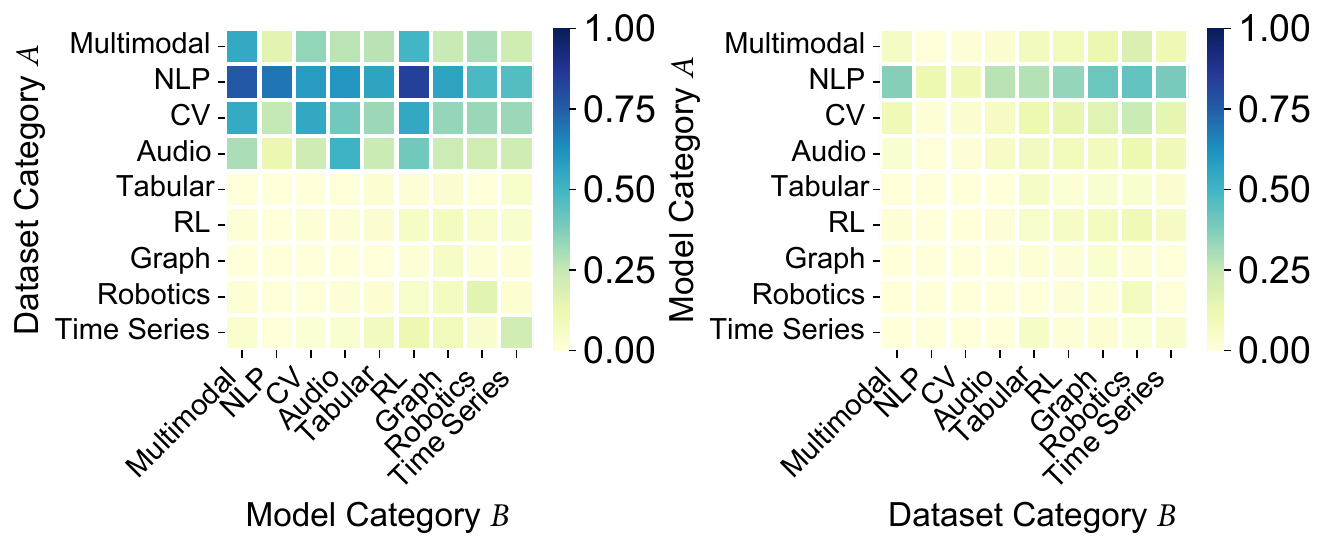}
        \caption{$P(A|B)$ for cross-type nodes.}
        \label{fig:colike-heatmap2}
    \end{subfigure}
    \caption{Conditional probability $P(A|B)$ of user co-likes.}
    \label{fig:colike-heatmap}
\end{figure}

\begin{figure}[t]
    \centering
    \includegraphics[width=\linewidth]{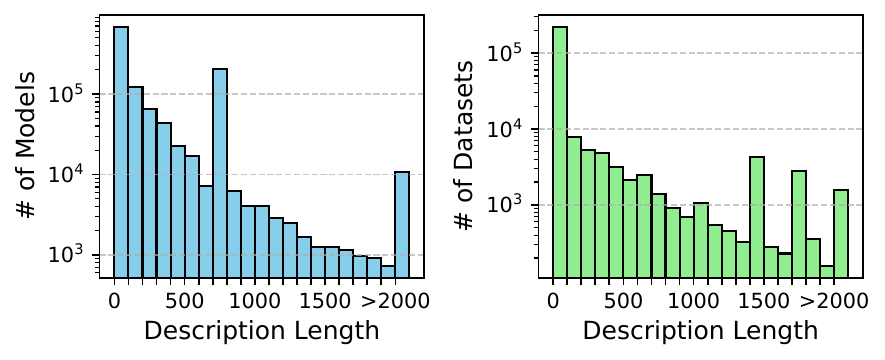}
    \caption{Description length of \texttt{Model} and \texttt{Dataset}.}
    \label{fig:description-length}
\end{figure}

\subsection{Statistics and Analysis}
\label{sec:kg-statistics}

\textbf{Distribution of Node and Relation Types.}
Figure~\ref{fig:schema} shows the distribution of node and relation types in $\mathsf{HuggingKG}$. Of the 2,614,270 nodes, \texttt{Model} makes up 46.0\%, followed by \texttt{User} (27.9\%) and \texttt{Space} (11.8\%). Smaller node types such as \texttt{Task} (0.0\%) and \texttt{Paper} (0.6\%) are important for identifying the characteristics of the resources. Among the 6,246,353 edges, user interactions such as \texttt{Like} (31.5\%) and \texttt{Publish} (28.8\%) dominate. Edges about model evolution, e.g., \texttt{Adapter} (2.5\%), \texttt{Finetune} (1.7\%), highlight the technical connections in the community. \emph{These patterns reflect a community-driven environment focused on user activity and model interoperability.}

\textbf{Resource Contextualization.}
To analyze patterns of user interest, we compute the conditional probabilities $P(A|B)$ of user co-likes for \texttt{Model} and \texttt{Dataset} across categories, as shown in Figure~\ref{fig:colike-heatmap}. NLP is the most popular research area, with users from other fields showing a strong interest in NLP resources. RL, in particular, is highly involved in NLP, likely due to advances in reinforcement learning with human feedback (RLHF) with large language models (LLMs). The multimodal community is also interested in NLP, CV, and audio resources. A comparison of Figures~\ref{fig:colike-heatmap1} and~\ref{fig:colike-heatmap2} shows clear differences in interest distribution, both within and between resource types. For example, a strong interest in NLP models is from users who like robotics datasets instead of NLP or RL, reflecting the intersection of embodied intelligence and LLMs. \emph{These patterns suggest that $\mathsf{HuggingKG}$ captures valuable information on user interests, useful for tasks like resource recommendation and trend prediction.}

\textbf{Textual Attributes.}
In $\mathsf{HuggingKG}$, the average length of the description is 270.2 words for \texttt{Model} and 134.1 words for \texttt{Dataset}, much longer than 8.1~words in~\cite{DEKR}. The longest \texttt{Model} description exceeds 2.5M words, and the longest \texttt{Dataset} description exceeds 400K words. As shown in Figure~\ref{fig:description-length}, the description length distribution exhibits a long tail, with 57.2\% of \texttt{Model} and 33.4\% of \texttt{Dataset} lacking descriptions. Peaks at 700--800 words for \texttt{Model} and 1500--1600 words for \texttt{Dataset} are mostly due to template usage. \emph{These patterns highlight the challenges of incomplete documentation and information overload, underlining the need for methods that improve metadata quality by combining textual and graph-based data.}



\section{$\mathsf{HuggingBench}$ Benchmark}
\label{sec:hugging-bench}
By combining search needs in open source resource communities with the structured information available in $\mathsf{HuggingKG}$, we identify three IR tasks in this domain: \textbf{resource recommendation} (Section~\ref{sec:bench-recommendation}), \textbf{task classification} (Section~\ref{sec:bench-classification}), and \textbf{model tracing} (Section~\ref{sec:bench-tracing}). We develop their test collections, which result in a novel benchmark, $\mathsf{HuggingBench}$.

\subsection{Resource Recommendation}
\label{sec:bench-recommendation}

\textbf{Application Scenario.}  
For ML practitioners, selecting an appropriate pre-trained model from thousands of options on platforms like Hugging Face is a significant challenge. For example, an NLP practitioner working on sentiment analysis requires a model tailored to their dataset and task. The \emph{resource recommendation} task addresses this by recommending models based on user interaction history. It enables practitioners to efficiently identify and deploy the most suitable model, reducing time and effort in resource discovery.

\textbf{Task Definition.}  
The resource recommendation task can be framed as a hybrid problem that integrates \emph{general collaborative filtering}, \emph{social recommendation}, and \emph{KG-based recommendation}. 

For general collaborative filtering, let $\mathcal{U} = \{u_1, u_2, ..., u_M\}$ represent the set of users and $\mathcal{I} = \{i_1, i_2, ..., i_N\}$ represent the set of items (e.g., \texttt{Model}, \texttt{Dataset}, and \texttt{Space}). The user-item interaction matrix $\mathbf{Y} = \{y_{ui} \mid u \in \mathcal{U}, i \in \mathcal{I}\}$ is captured from \texttt{Like} edges, where $y_{ui} = 1$ if \texttt{User} $u$ liked item $i$, and $y_{ui} = 0$ otherwise. This matrix serves as the foundation for general recommendation systems. 

Social recommendation is facilitated by using social relationships between users represented by a graph $\mathcal{S} = \{s_{uv} \mid u, v \in \mathcal{U}\}$ where $s_{uv} = 1$ indicates a \texttt{Follow} edge between \texttt{User} nodes $u$ and $v$. 

KG-based recommendation draws on structured external knowledge from the $\mathsf{HuggingKG}$ knowledge graph, denoted $\mathcal{G} = (\mathcal{V}, \mathcal{E})$, which encodes external entities (e.g.,~\texttt{Paper}, \texttt{Collection}) and their interrelations as edges (a.k.a. triples in this context).

The objective of the recommendation task is to learn a prediction function $\hat{y}_{ui} = \mathcal{F}(u, i)$, where $\hat{y}_{ui}$ is the predicted probability that user $u$ interacts with item $i$. The goal is to rank the items for each user based on these predicted probabilities, prioritizing those with the highest likelihood of interaction to ensure that the most relevant items appear at the top of the recommendation list.

\begin{table}
    \centering
    \small
    \caption{Test collection for resource recommendation. Avg. means the average number of interactions per user.}
    \begin{tabular}{lrrrr}
    \toprule
     & \textbf{\#Users} & \textbf{\#Items} & \textbf{\#Interactions} & \textbf{Avg.} \\
    \midrule
    \texttt{Model} & 29,720 & 16,200 & 667,365 & 22.46 \\
    \texttt{Dataset} & 5,072 & 3,634 & 100,561 & 19.83 \\
    \texttt{Space} & 14,171 & 5,246 & 297,294 & 20.98 \\
    \midrule
    Total & 38,624 & 25,080 & 1,065,220 & 27.58 \\
    \bottomrule
    \end{tabular}
    \label{tab:recommendation-dataset}
\end{table}

\begin{table}
\centering
    \small
    \caption{External KGs for KG-based recommendation.}
    \begin{tabular}{lrrrr}
    \toprule
   & 2hop & 1hop & Homo & Publish \\
    \midrule
    \textbf{\#Nodes} & 1,242,578 & 462,213 & 25,080 & 31,754 \\
    \textbf{\#Edges (Triples)} & 3,063,081 & 1,180,911 & 28,346 & 25,078 \\
    \textbf{\#Relations} & 27 & 27 & 7 & 6 \\
    \bottomrule
    \end{tabular}
    \label{tab:recommendation-subgraphs}
\end{table}

\textbf{Dataset Construction.}
Based on $\mathsf{HuggingKG}$, we extract a bipartite graph representing the \texttt{Like} edges between \texttt{User} nodes and three types of item nodes (\texttt{Model}, \texttt{Dataset}, and \texttt{Space}). Following standard practices in recommendation tasks, we derive a 5-core subgraph~(where each node has at least 5 edges to other nodes) from this bipartite graph. Subsequently, we partition the liked-item list for each user into training, validation, and test sets with a split of 60\%, 20\%, and 20\%, respectively. Table~\ref{tab:recommendation-dataset} shows statistics of the resource recommendation test collection.

To support social recommendation, we introduce 84,913 \texttt{Follow} edges between \texttt{User} nodes in the test collection as external \textbf{Social} information. To support KG-based recommendation, we consider three types of subgraphs of $\mathsf{HuggingKG}$ as external KGs: \textbf{1hop}/\textbf{2hop} are 1-step/2-step neighborhood subgraphs from item-aligned KG nodes, \textbf{Homo} is the subgraph induced from all item nodes, and \textbf{Publish} is the relation-specific subgraph induced from the \texttt{Publish} edges. Table~\ref{tab:recommendation-subgraphs} shows statistics of these external KGs.

\subsection{Task Classification}
\label{sec:bench-classification}

\textbf{Application Scenario.}
Platform curators face the challenge of organizing and tagging models and datasets to improve searchability and usability. For example, a newly uploaded model without clear task annotations becomes difficult for users to discover. The \emph{task classification} task automates this process by analyzing the metadata and structure information from the platforms to classify the models into relevant tasks (e.g.,~``text classification'' or ``named entity recognition''). This ensures proper categorization and improves accessibility for users seeking specific features.

\textbf{Task Definition.}
The task of classifying \texttt{Model} and \texttt{Dataset} nodes by their associated \texttt{Task}~(according to \texttt{Defined For} edges) is framed as a \emph{multi-label attributed node classification} problem, where each instance can be assigned multiple labels from a set $\mathcal{L} = \{l_1, l_2, \dots, l_K\}$. The input consists of a graph $\mathcal{G} = (\mathcal{V}, \mathcal{E})$, where the nodes $\mathcal{V}$ represent \texttt{Model} and \texttt{Dataset} nodes, and the edges $\mathcal{E}$ captured include~\texttt{Finetune} edges between \texttt{Model} nodes, \texttt{Trained Or Finetuned On} edges between \texttt{Model} nodes and \texttt{Dataset} nodes, etc. Additionally, a feature matrix $\mathbf{X} \in \mathbb{R}^{|\mathcal{V}| \times d}$ encodes node features such as textual descriptions or metadata. The output for each node $v \in \mathcal{V}$ is a binary vector $\mathbf{y}_v \in \{0, 1\}^K$, where $y_{v,k} = 1$ indicates an association with task $t_k$. The goal is to learn a function $\hat{\mathbf{y}}_{v} = \mathcal{F} (\mathcal{G}, \mathbf{X})$ that accurately predicts task labels, using both graph structure and textual information, while optimizing precision and recall across all labels.

\begin{table}
    \centering
    \caption{Test collection for task classification.}
    \resizebox{\linewidth}{!}{
    \begin{tabular}{lrrrrrr}
        \toprule
        & \textbf{\#Nodes} & \textbf{\#Edges} & \textbf{\#Labels} & \textbf{\#Train} & \textbf{\#Valid} & \textbf{\#Test} \\
        \midrule
        \texttt{Model} & 145,466 & 131,274 & 50 & 103,276 & 19,167 & 27,974 \\
        \texttt{Dataset} & 6,969 & - & 48 & 6,792 & 1,501 & 708 \\
        \midrule
        Total & 152,435 & 166,199 & 52 & 110,068 & 20,668 & 28,682 \\
        \bottomrule
    \end{tabular}
    }
    \label{tab:classification-dataset}
\end{table}

\textbf{Dataset Construction.}
We first select the \texttt{Model} and \texttt{Dataset} nodes in the graph that have associated task labels, based on the \texttt{Defined For} edges between the \texttt{Model}/\texttt{Dataset} nodes and the \texttt{Task} nodes in $\mathsf{HuggingKG}$. We then add all edges between these nodes and remove any isolated nodes. To better align the task with the scenario of helping the community website automatically predict the task types of newly created models or datasets, we divide the data into training, validation, and test sets based on the creation dates of \texttt{Model} and \texttt{Dataset}. Due to the faster release rate of models compared to datasets, we set a longer time range for \texttt{Dataset} than for \texttt{Model}. Specifically, we select \texttt{Model} nodes created between 2024-08-15 and 2024-10-15, and \texttt{Dataset} nodes created between 2024-04-15 and 2024-08-15 for the validation set. The test set comprises \texttt{Model} nodes created between 2024-10-15 and 2024-12-15, as well as \texttt{Dataset} nodes created between 2024-08-15 and 2024-12-15. The remaining \texttt{Model} and \texttt{Dataset} nodes, released before these periods, are used as the training set. Table~\ref{tab:classification-dataset} shows statistics of the task classification test collection.

\subsection{Model Tracing}
\label{sec:bench-tracing}

\textbf{Application Scenario.}  
Researchers often need to investigate the lineage and dependencies of model architectures, such as understanding the evolution of GPT-3. The \emph{model tracing} task facilitates this by tracing the base model of a model, identifying base models (e.g.,~GPT-2) and related variants, along with their associated datasets and tasks. This capability supports in-depth analysis of model development histories and their connections within the community, aiding reproducibility and innovation.

\textbf{Task Definition.}
The model tracing task can be formally defined as a specialized \emph{link prediction} problem within a heterogeneous graph $\mathcal{G} = (\mathcal{V}, \mathcal{E})$, where $\mathcal{V}$ represents the set of nodes, including \texttt{Model} and other nodes (e.g.,~\texttt{Dataset}, \texttt{Space}), and $\mathcal{E}$ represents the set of edges. Each edge is represented as a triple $(h, r, t)$, where $h \in \mathcal{V}$ is the head node, $r \in \mathcal{R}$ is the relation (i.e.,~\texttt{Adapter}, \texttt{Finetune}, \texttt{Merge} or \texttt{Quantize} between \texttt{Model} nodes), and $t \in \mathcal{V}$ is the tail node. Unlike general link prediction tasks, this task specifically predicts the reverse relation between \texttt{Model} nodes: given a relation~$r$ and a tail node~$t$, the goal is to predict the corresponding head node~$h$ that completes the triple $(h, r, t)$. The output of the task is a probability distribution $\mathbf{y} \in [0, 1]^{|\mathcal{V}|}$ on all candidate head nodes, with~$y_h$ indicating the probability that node~$h$ is the correct match for the triple $(h, r, t)$. The objective is to learn a function $\hat{\mathbf{y}}_{(r,t)}=\mathcal{F}(\mathcal{G}, r, t)$ that maximizes the likelihood of correctly identifying the true head node $h^*$ for the triple $(h^*, r, t)$.

\begin{table}
    \small
    \centering
    \caption{Test collection for model tracing.}
    \begin{tabular}{lrrr}
    \toprule
          & \textbf{\#Train} & \textbf{\#Valid} & \textbf{\#Test} \\
        \midrule
        \texttt{Adapter} & 565 & 65 & 80 \\
        \texttt{Finetune} & 15,639 & 1,944 & 1,935 \\
        \texttt{Merge} & 138 & 24 & 17 \\
        \texttt{Quantize} & 178 & 16 & 23 \\
        \midrule
        Total & 16,520 & 2,049 & 2,055 \\
    \bottomrule
    \end{tabular}
    \label{tab:tracing-dataset}
\end{table}

\textbf{Dataset Construction.}
Due to the large size of the complete $\mathsf{HuggingKG}$, we extract a subgraph to construct the test collection for model tracing. We select the \texttt{Model} nodes associated with the task type ``text classification''. Next, we identify the triples where these nodes appear as the head or tail node, incorporating the other node in the triple into the node set~$\mathcal{V}$. We then include all edges between these nodes to form the edge set~$\mathcal{E}$. The resulting subgraph consists of 121,404 nodes, 339,429 edges, and 30 relations.

We subsequently partition the inter-\texttt{Model} edges into training, validation, and test sets with a split of 80\%, 10\%, and 10\%, respectively. Table~\ref{tab:tracing-dataset} shows statistics of the model tracing test collection.

\begin{table*}[]
  \centering
    \small
  \caption{Evaluation results of resource recommendation.}
    \begin{tabular}{lccccccccc}
    \toprule
    \textbf{Method} & \textbf{KG} & \textbf{Recall@5} & \textbf{Recall@10} & \textbf{Recall@20} & \textbf{Recall@40} & \textbf{NDCG@5} & \textbf{NDCG@10} & \textbf{NDCG@20} & \textbf{NDCG@40} \\
    \midrule
    \rowcolor{cyan!40} \multicolumn{10}{l}{\emph{General Collaborative Filtering}}  \\
    \textbf{LightGCN} & - & 0.0856 & 0.1301 & 0.1932 & 0.2759 & 0.0868 & 0.1003 & 0.1192 & 0.1413 \\
    \textbf{HCCF} & - & 0.0834 & 0.1254 & 0.1820 & 0.2504 & 0.0847 & 0.0975 & 0.1143 & 0.1328 \\
    \textbf{SimGCL} & - & 0.0999 & 0.1515 & 0.2186 & 0.3010 & 0.0998 & 0.1158 & 0.1358 & 0.1581 \\
    \textbf{LightGCL} & - & \textbf{0.1033} & \textbf{0.1558} & \textbf{0.2228} & 0.3017 & \textbf{0.1035} & \textbf{0.1198} & \textbf{0.1398} & \textbf{0.1611} \\
    \textbf{AutoCF} & - & 0.1003 & 0.1530 & 0.2190 & \textbf{0.3039} & 0.1012 & 0.1174 & 0.1371 & 0.1598 \\
    \textbf{DCCF} & - & 0.0985 & 0.1493 & 0.2167 & 0.3003 & 0.0983 & 0.1142 & 0.1343 & 0.1567 \\
    \rowcolor{yellow!40} \multicolumn{10}{l}{\emph{Social Recommendation}}  \\
    \textbf{MHCN} & \multirow{2}{*}{Social} & \textbf{0.0979} & \textbf{0.1490} & \textbf{0.2162} & \textbf{0.3007} & \textbf{0.0998} & \textbf{0.1154} & \textbf{0.1353} & \textbf{0.1579} \\
    \textbf{DSL} & & 0.0932 & 0.1425 & 0.2123 & 0.2986 & 0.0948 & 0.1099 & 0.1307 & 0.1538 \\
    \rowcolor{pink!40} \multicolumn{10}{l}{\emph{KG-Based Recommendation}}  \\
    \textbf{KGIN} & \multirow{3}{*}{1hop} & 0.0001 & 0.0004 & 0.0010 & 0.0017 & 0.0002 & 0.0003 & 0.0005 & 0.0007 \\
    \textbf{KGCL} &  & 0.0993 & 0.1490 & 0.2135 & 0.2918 & 0.1009 & 0.1160 & 0.1351 & 0.1563 \\
    \textbf{KGRec} & & 0.0558 & 0.0897 & 0.1395 & 0.2076 & 0.0575 & 0.0681 & 0.0832 & 0.1014 \\
    \midrule
    \textbf{KGIN} & \multirow{3}{*}{2hop} & 0.0002 & 0.0004 & 0.0008 & 0.0016 & 0.0003 & 0.0004 & 0.0005 & 0.0007 \\
    \textbf{KGCL} &  & 0.1007 & 0.1510 & 0.2165 & 0.2959 & 0.1016 & 0.1170 & 0.1364 & 0.1579 \\
    \textbf{KGRec} & & 0.0597 & 0.0941 & 0.1423 & 0.2122 & 0.0625 & 0.0729 & 0.0872 & 0.1057 \\
    \midrule
    \textbf{KGIN} & \multirow{3}{*}{Homo} & 0.0061 & 0.0096 & 0.0146 & 0.0219 & 0.0065 & 0.0076 & 0.0091 & 0.0111 \\
    \textbf{KGCL} &  & \textbf{0.1054} & \textbf{0.1578} & \textbf{0.2237} & \textbf{0.3059} & \textbf{0.1058} & \textbf{0.1220} & \textbf{0.1416} & \textbf{0.1637}  \\
    \textbf{KGRec} & & 0.0628 & 0.0985 & 0.1476 & 0.2106 & 0.0638 & 0.0751 & 0.0898 & 0.1067 \\
    \midrule
    \textbf{KGIN} & \multirow{3}{*}{Publish} & 0.0002 & 0.0003 & 0.0007 & 0.0016 & 0.0002 & 0.0003 & 0.0004 & 0.0007 \\
    \textbf{KGCL} & & 0.1036 & 0.1543 & 0.2205 & 0.3011 & 0.1038 & 0.1195 & 0.1392 & 0.1609 \\
    \textbf{KGRec} & & 0.0609 & 0.0941 & 0.1385 & 0.2002 & 0.0636 & 0.0734 & 0.0863 & 0.1027 \\
    \bottomrule
    \end{tabular}
    \label{tab:result-recommendation}
\end{table*}

\section{Evaluation}
\label{sec:evaluations}



We evaluate with $\mathsf{HuggingBench}$. All experiments are conducted on a server with four NVIDIA Tesla V100 SXM2 32 GB GPUs.

\subsection{Resource Recommendation}
\label{sec:evaluations-recommendation}

\textbf{Evaluation Metrics.}
Following standard practices in recommender systems~\cite{SSLRec, LightGCN}, we use Recall@$K$ and NDCG@$K$ as evaluation metrics to evaluate the ranked list of recommended items. The value of $K$ is set to 5, 10, 20, and 40 for the evaluation.

\textbf{Baselines.}
Following common practice in recommendation~\cite{SSLRec, RLMRec}, we select six representative \emph{general collaborative filtering} methods, including four graph-based methods: 
 \textbf{LightGCN}~\cite{LightGCN}, \textbf{HCCF}~\cite{HCCF}, \textbf{SimGCL}~\cite{SimGCL}, and \textbf{LightGCL}~\cite{LightGCL} and two representation learning methods: \textbf{AutoCF}~\cite{AutoCF} and \textbf{DCCF}~\cite{DCCF}. These methods rely on the bipartite user-item graph for interaction modeling.
We also adopt two state-of-the-art \emph{social recommendation} methods: \textbf{MHCN}~\cite{MHCN} and \textbf{DSL}~\cite{DSL}. These methods introduce the social graph to capture richer user preferences. 
In addition, we employ three \emph{KG-based recommendation} methods: \textbf{KGIN}~\cite{KGIN}, \textbf{KGCL}~\cite{KGCL}, and \textbf{KGRec}~\cite{KGRec}. These methods utilize a unified heterogeneous structure that aligns items in the bipartite graph with nodes from external KGs. We use SSLRec\footnote{\url{https://github.com/HKUDS/SSLRec}}~\cite{SSLRec} to implement the above methods.

\textbf{Implementation Details.}
For all baseline models, the representation dimension is set to 64. Each model is trained for up to 100 epochs, with a fixed batch size of 4,096 and early stopping based on MRR@5 on the validation set. Validation is performed every 3 epochs, and the patience of early stopping is set to 5.  We perform a grid search to select the optimal learning rate from \{1e-3, 1e-4, 1e-5\} and the number of GNN layers from \{2, 3\}.

\textbf{Evaluation Results.}
As shown in Table~\ref{tab:result-recommendation}, among the general collaborative filtering methods, LightGCL achieves the best performance, suggesting that \emph{interaction graph augmentation through singular value decomposition provides a strong baseline}.

For social recommendation, MHCN outperforms DSL by +4.80\% in Recall@$5$, showing the effectiveness of multi-channel hypergraph convolution and self-supervised learning. However, it surpasses only two of the six general collaborative filtering methods, indicating that \emph{adding social data does not secure improvement.}

For KG-based recommendation, KGCL consistently outperforms others in all subgraph types, highlighting the benefits of KG augmentation and contrastive learning. Homo subgraph yields the best results, suggesting that \emph{item-related nodes provide high-quality context}. In sparse graphs, KGCL is robust due to the use of contrastive learning with well-defined positive and negative pairs, which improves the quality of learned representations. In contrast, KGIN's performance is highly sensitive to its negative sampling strategy and hyperparameter setting, often instable in sparse or cold-start scenarios. KGRec relies heavily on the quality of entity representations and performs moderately when the graph density is low.

Moreover, KG-based methods exhibit higher variance, with Recall@$5$ having a standard deviation of $0.0431$, compared to $0.0084$ and $0.0033$ for social and collaborative filtering methods. This suggests that \emph{applying KG-based methods requires careful selection and tuning of the specific approach based on data characteristics such as graph density and relation sparsity.}


\textbf{Comparison with Other Benchmarks.}
To investigate the underwhelming performance of social recommendation methods, we compare the social relation statistics of our test collection with those of the LastFM, Douban and Yelp datasets where MHCN has demonstrated strong performance~\cite{MHCN}. LastFM includes $1,892$ users with $25,434$ relations, Douban has $2,848$ users with $35,770$ relations, and Yelp contains $19,535$ users with $864,157$ relations, all of which are considerably denser than our test collection, which consists of $38,624$ users and $84,913$ relations. MHCN relies on dense connections to form meaningful hyperedges, facilitate contrastive learning, and effectively propagate signals. Similarly, despite KGIN's strong performance on the Amazon-Book, LastFM, and Alibaba datasets~\cite{KGIN}, it performs the worst on our test collection. A possible factor is the KG sparsity: Amazon-Book has $2,557,746$ triples and $88,572$ nodes, LastFM has $464,567$ triples and $58,266$ nodes, and Alibaba has $279,155$ triples and $59,156$ nodes, while our dataset is significantly sparser (Table~\ref{tab:recommendation-subgraphs}). KGIN relies on dense multi-hop paths ($\ge3$ hops) for user-item semantics, which are often missing in sparse KGs. Compared with the aforementioned benchmarks, the metrics of various baselines on our test collection are generally lower, reflecting the difficulty of this recommendation task. \emph{In summary, because our test collection provides more domain-specific external relations, which are relatively sparse and specialized compared to relations aligned from huge KGs such as Wikidata, it provides new challenges and benchmarks for recommendation methods that use special additional information for domain adaptation.}

\subsection{Task Classification}
\label{sec:evaluations-classification}

\textbf{Evaluation Metrics.}
Following standard practices in multi-label node classification~\cite{CogDL}, we use Micro-F1.

\textbf{Baselines.}
We select nine representative \emph{GNN-based} methods for node classification. \textbf{GCN}~\cite{GCN}, \textbf{GAT}~\cite{GAT}, and \textbf{GraphSAGE}~\cite{GraphSAGE} establish fundamental architectures with different aggregation schemes. Memory and computation optimization approaches are represented by \textbf{GraphSAINT}~\cite{GraphSAGE}, \textbf{RevGCN}~\cite{RevGNN}, and \textbf{RevGAT}~\cite{RevGNN}. The remaining methods (\textbf{APPNP}~\cite{APPNP}, \textbf{GRAND}~\cite{GRAND}, \textbf{GCNII}~\cite{GCNII}) focus on addressing specific challenges like label propagation, robustness, and over-smoothing in deep GNNs.
We use CogDL~\cite{CogDL}\footnote{\url{https://github.com/THUDM/cogdl}} to implement the above methods.

For node feature initialization, we try the following settings:

\begin{itemize}[leftmargin=*]
    \item \textbf{Binary}: Binary one-dimensional vectors for distinguishing between \texttt{Model} (0) and \texttt{Dataset} (1).
    \item Pre-trained Text Embeddings: Embeddings of node description attributes generated using base versions of \textbf{BERT}\footnote{\url{https://huggingface.co/google-bert/bert-base-uncased}} and \textbf{BGE}.\footnote{\url{https://huggingface.co/BAAI/bge-base-en-v1.5}}
    \item Finetuned Text Embeddings: Node description attribute embeddings derived from the aforementioned models after finetuning for 1 epoch. We perform a grid search to select the optimal learning rate from \{1e-4, 5e-5, 1e-5\} and batch size from \{8, 16\}. The resulting models are denoted as \textbf{BERT (ft)} and \textbf{BGE (ft)}.
\end{itemize}

\textbf{Implementation Details.}
The hidden size dimension is set to 1,024 for all baselines based on GNN, except GAT and RevGAT, for which it is set to 256 to accommodate GPU memory limitations. Each GNN model is trained for up to 500 epochs, with early stopping based on the Micro-F1 score on the validation set. The patience of validation early stopping is set to 100 epochs. For the training process, the batch size is set at 4,096. We perform a grid search to select the optimal learning rate from \{1e-3, 1e-4, 1e-5\}, weight decay from \{0, 1e-5, 1e-4\}, and the number of GNN layers from \{2, 3, 4\}.

\begin{table}
  \centering
    \small
  \caption{Evaluation results (Micro-F1) of task classification.}
  \begin{tabular}{lccccc}
    \toprule
    \textbf{Method} & \textbf{Binary} & \textbf{BERT} & \textbf{BERT (ft)} & \textbf{BGE} & \textbf{BGE (ft)} \\
    \midrule
    \textbf{GCN} & 0.0662 & \textbf{0.7620} & 0.8291 & 0.7411 & 0.8522 \\
    \textbf{GAT} & 0.0390 & 0.5105 & 0.8125 & 0.5444 & 0.8261 \\
    \textbf{GRAND} & 0.1228 & 0.1297 & 0.6089 & 0.2646 & 0.4532 \\
    \textbf{GraphSAGE} & \textbf{0.1800} & 0.5341 & 0.8845 & \textbf{0.8199} & \textbf{0.8830} \\
    \textbf{APPNP} & 0.0448 & 0.7297 & 0.8304 & 0.7571 & 0.8419 \\
    \textbf{GCNII} & 0.1149 & 0.6456 & 0.8836 & 0.7779 & 0.8802 \\
    \textbf{GraphSAINT} & 0.0579 & 0.2703 & 0.8342 & 0.0540 & 0.8251 \\
    \textbf{RevGCN} & 0.1071 & 0.6763 & \textbf{0.8851} & 0.8039 & 0.8770 \\
    \textbf{RevGAT} & 0.0335 & 0.7412 & 0.8849 & 0.7569 & 0.8716 \\
    \bottomrule
  \end{tabular}
  \label{tab:results-classification}
\end{table}

\textbf{Evaluation Results.}
As shown in Table~\ref{tab:results-classification}, all models demonstrate limited performance with binary features, with GraphSAGE achieving the best score of only 0.1800. When using pre-trained embeddings, we observe significant improvements across most models, with GCN achieving 0.7620 with BERT embeddings and GraphSAGE reaching 0.8199 with BGE embeddings. For finetuned embeddings, the performance improves substantially, with RevGCN achieving 0.8851 and GraphSAGE reaching 0.8830 with finetuned BERT and BGE respectively, followed closely by GCNII. The experimental results clearly demonstrate that using pre-trained text embeddings as initial features significantly enhances model performance, with an average improvement of 0.5391 in the Micro-F1 score. Furthermore, finetuning the embeddings for task-specific feature learning consistently brings additional gains across all methods and embeddings, with an average improvement of 0.2543 and 0.1871 over BERT and BGE pre-trained embeddings, respectively, \emph{suggesting that incorporating domain knowledge through finetuning is important for improving performance in this task}.

We notice an interesting pattern where simpler architectures like GraphSAGE outperform sophisticated models such as GCNII and RevGAT, particularly with BGE embeddings. \emph{This suggests that complex architectural designs might not always be beneficial when working with high-quality pre-trained embeddings, as they may introduce noise or over-smoothing in the feature space.}

\textbf{Comparison with Other Benchmarks.}
We find that sampling-based models such as GRAND and GraphSAINT exhibit unexpectedly poor performance with pre-trained embeddings (0.1297 and 0.2703 with BERT, respectively), despite their effectiveness~\cite{CogDL} in other multi-label node classification datasets such as ogbn-arxiv~\cite{obgn-arxiv}. This performance degradation might be attributed to their sampling strategies potentially disrupting the semantic relations encoded in the pre-trained embeddings. Compared with other commonly used benchmarks, the metrics of our test collection of various baselines are at a medium level~\cite{CogDL}. \emph{In summary, for such a new domain-specific classification task, it not only poses new challenges to text embedding models, but also generates new issues worthy of consideration for the GNN-based node classification methods.}

\begin{table}
  \centering
    \small
  \caption{Evaluation results of model tracing.}
  \begin{tabular}{lccccc}
  \toprule
    \textbf{Method} & \textbf{MRR} & \textbf{Hit@1} & \textbf{Hit@3} & \textbf{Hit@5} & \textbf{Hit@10} \\
  \midrule
  \rowcolor{yellow!40} \multicolumn{6}{l}{\emph{Supervised}} \\
  \textbf{RESCAL} & 0.2694 & 0.2380 & 0.2667 & 0.2929 & 0.3470 \\
  \textbf{TransE} & \textbf{0.5589} & \textbf{0.4496} & \textbf{0.6321} & \textbf{0.6973} & \textbf{0.7562} \\
  \textbf{DistMult} & 0.2050 & 0.1421 & 0.2321 & 0.2735 & 0.3324 \\ 
  \textbf{ComplEx} & 0.1807 & 0.1109 & 0.2122 & 0.2599 & 0.3066 \\
  \textbf{ConvE} & 0.4739 & 0.3766 & 0.5119 & 0.5903 & 0.6735 \\
  \textbf{RotatE} & 0.5317 & 0.4195 & 0.6029 & 0.6803 & 0.7392 \\
  \textbf{HittER} & 0.3678 & 0.2900 & 0.4078 & 0.4657 & 0.5314 \\
  \rowcolor{pink!40} \multicolumn{6}{l}{\emph{Unsupervised}} \\
  \textbf{ULTRA} & 0.3373 & 0.1440 & \textbf{0.4803} & \textbf{0.5309} & \textbf{0.6672} \\
  \textbf{KG-ICL} & \textbf{0.4008} & \textbf{0.3354} & 0.3792 & 0.4854 & 0.5938 \\
  \bottomrule
  \end{tabular}
  \label{tab:result-tracing}
\end{table}

\subsection{Model Tracing}
\label{sec:evaluations-tracing}

\textbf{Evaluation Metrics.} 
Following standard practices in link prediction~\cite{KG-ICL, RotatE}, we use Mean Reciprocal Rank~(MRR) and Hit@$K$ as evaluation metrics for the model tracing experiments, with $K$ set to 1, 3, 5, and 10. As mentioned in Section~\ref{sec:bench-tracing}, we only evaluate the results considering the given \texttt{Model} node as the tail node $t$ and the given relation $r$ to predict the head node $h$.

\textbf{Baselines.}
According to common practice in the field of link prediction~\cite{LibKGE, ULTRA, KG-ICL}, we select seven \emph{supervised} methods, including five embedding-based methods: \textbf{RESCAL}~\cite{RESCAL}, \textbf{TransE}~\cite{TransE}, \textbf{DistMult}~\cite{DistMult}, \textbf{ComplEx}~\cite{ComplEX}, and \textbf{RotatE}~\cite{RotatE} and two deep neural network-based methods: CNN-based \textbf{ConvE}~\cite{ConvE} and transformer-based \textbf{HittER}~\cite{HittER}. Among these, \textbf{TransE} is one of the most classic methods, which simply represents nodes as continuous vectors in the embedding space and represents the relations between nodes as translation vectors from the head node to the tail node. We also adopt two recent \emph{unsupervised} KG foundation models, \textbf{ULTRA}~\cite{ULTRA} and \textbf{KG-ICL}~\cite{KG-ICL}. These models represent relations using relation graphs and prompt graphs, respectively, which encode nodes and their scores. Since they do not rely on specific learnable representation vectors for nodes or relations, they can directly utilize frozen pre-trained models on our test collection.
We use LibKGE~\cite{LibKGE}\footnote{\url{https://github.com/uma-pi1/kge}} to implement supervised methods and use the official code\footnote{\url{https://github.com/DeepGraphLearning/ULTRA}}\footnote{\url{https://github.com/nju-websoft/KG-ICL}} of the two unsupervised models.

\textbf{Implementation Details.}
For all the baseline methods, the representation dimension is set to 64. Each baseline is trained for up to 100 epochs, with early stopping based on the MRR score on the validation set. Validation is performed every 3 epochs, and the patience of early stopping is set to 5. For training process, the batch size is fixed at 256. We perform a grid search to select the optimal learning rate from \{1e-2, 1e-3, 1e-4, 1e-5\} and the number of GNN layers from \{2, 3\}.

\textbf{Evaluation Results.}
The results for the model tracing task are presented in Table~\ref{tab:result-tracing}. Supervised models, particularly TransE, outperform all other methods, achieving the highest scores in all metrics. In contrast, bilinear models like RESCAL and DistMult, and neural models like ConvE, underperform due to their reliance on multiplicative interactions or convolutions, which fail to capture critical dependencies. \emph{In conclusion, simpler geometric transformations are better suited for this task.}

For unsupervised models, KG-ICL excels in MRR and Hit@1, while ULTRA performs better in Hit@3/5/10 but struggles with accuracy in top-ranked predictions. This difference can be attributed to that ULTRA performs message passing across the entire graph, while KG-ICL initiates message passing from the head node, expanding one hop at each layer, which allows it to focus more on nearby nodes. As a result, KG-ICL achieves a higher Hit@1 score. \emph{In conclusion, unsupervised models exhibit a good trade-off between precision and recall for this task.}

\textbf{Comparison with Other Benchmarks.}
In contrast to the results of widely used KG link prediction benchmarks such as FB15k-237~\cite{FB15k-237} and WN18RR~\cite{WN18RR}, TransE performs the best among supervised methods. Although TransE is a classic and effective method, its linear modeling struggles with one-to-many relations (where a single head node and a relation can correspond to multiple tail nodes)~\cite{RotatE}. Consequently, many subsequent models focus on addressing this issue. We analyze the test sets of FB15k-237, WN18RR and our dataset, counting the number of answers corresponding to each (head node, relation) pair. The results show that FB15k-237 has an average of 399.82 answers, WN18RR has 24.57, while our dataset has only 1.04. It indicates that the answers in our test set are nearly unique, making TransE sufficient for our needs. More complex methods tend to focus on distinguishing similar nodes at the many end of one-to-many relations, resulting in suboptimal performance. Compared to the metrics of these baselines on FB15k-237 and WN18RR, the metrics on our test collection are generally lower, suggesting that our test collection is overall more challenging. \emph{In summary, the unique distribution of relations in our test collection makes model tracing difficult for existing methods and raises new requirements for approaches designed for this task.}


\section{Conclusion}
\label{sec:conclusion} 

\textbf{Predicted Impact.}  
$\mathsf{HuggingKG}$ and $\mathsf{HuggingBench}$ will enable impactful IR research activities, particularly in advancing resource recommendation and automatic management within open source communities. These resources not only advance well-established research areas (e.g.,~recommendation, classification, and KG), but also foster emerging domains related to LLM. For example, by introducing $\mathsf{HuggingKG}$ into the model selection step of methods such as HuggingGPT~\cite{HuggingGPT}, it has the potential to improve the integration of AI tools for LLMs. Given their foundational nature, these resources are designed to remain relevant and useful over an extended period. To ensure long-term sustainability, we plan to release an updated version of $\mathsf{HuggingKG}$ periodically and have provided open source code for KG and benchmark construction, allowing the community to maintain and customize the resources for various use cases. The anticipated research user community spans multiple disciplines, including IR, ML, data mining, and software engineering, with a substantial base of current users. This community is expected to grow significantly in the coming years as the demand for structured knowledge and robust evaluation frameworks increases, driven by advances in the communities of LLM and AI tools. Taken together, these contributions position $\mathsf{HuggingKG}$ and $\mathsf{HuggingBench}$ as enduring assets for the academic and industrial communities.

\textbf{Limitations and Future Work.}
Our resources have the following limitations to be addressed in future work. First, the data is currently limited to Hugging Face, which restricts the diversity of entities and relations. In future work, we plan to expand $\mathsf{HuggingKG}$ to include additional resource platforms such as GitHub and Kaggle to introduce more types of cross-platform entities and relations. This expansion will enable applications such as cross-domain entity alignment. Second, the test collections in $\mathsf{HuggingBench}$ are automatically generated from $\mathsf{HuggingKG}$ and lack tasks that require manual annotation, such as question answering (QA), limiting the breadth of the benchmark. To address this, we will focus on designing more advanced LLM agents to support annotation for IR tasks including QA and retrieval, or combining graph-based methods and manual annotations for richer benchmarks.

\begin{acks}
This work was supported by the Postgraduate Research \& Practice Innovation Program of Jiangsu Province.
\end{acks}

\clearpage

\bibliographystyle{ACM-Reference-Format}
\balance
\bibliography{main}


\begin{thebibliography}{51}


\ifx \showCODEN    \undefined \def \showCODEN     #1{\unskip}     \fi
\ifx \showISBNx    \undefined \def \showISBNx     #1{\unskip}     \fi
\ifx \showISBNxiii \undefined \def \showISBNxiii  #1{\unskip}     \fi
\ifx \showISSN     \undefined \def \showISSN      #1{\unskip}     \fi
\ifx \showLCCN     \undefined \def \showLCCN      #1{\unskip}     \fi
\ifx \shownote     \undefined \def \shownote      #1{#1}          \fi
\ifx \showarticletitle \undefined \def \showarticletitle #1{#1}   \fi
\ifx \showURL      \undefined \def \showURL       {\relax}        \fi
\providecommand\bibfield[2]{#2}
\providecommand\bibinfo[2]{#2}
\providecommand\natexlab[1]{#1}
\providecommand\showeprint[2][]{arXiv:#2}

\bibitem[Bai et~al\mbox{.}(2024)]%
        {issue-PR-link-prediction}
\bibfield{author}{\bibinfo{person}{Shuotong Bai}, \bibinfo{person}{Huaxiao Liu}, \bibinfo{person}{Enyan Dai}, {and} \bibinfo{person}{Lei Liu}.} \bibinfo{year}{2024}\natexlab{}.
\newblock \showarticletitle{Improving Issue-PR Link Prediction via Knowledge-Aware Heterogeneous Graph Learning}.
\newblock \bibinfo{journal}{\emph{{IEEE} Trans. Software Eng.}} \bibinfo{volume}{50}, \bibinfo{number}{7} (\bibinfo{year}{2024}), \bibinfo{pages}{1901--1920}.
\newblock
\href{https://doi.org/10.1109/TSE.2024.3408448}{doi:\nolinkurl{10.1109/TSE.2024.3408448}}


\bibitem[Bordes et~al\mbox{.}(2013)]%
        {TransE}
\bibfield{author}{\bibinfo{person}{Antoine Bordes}, \bibinfo{person}{Nicolas Usunier}, \bibinfo{person}{Alberto Garc{\'{\i}}a{-}Dur{\'{a}}n}, \bibinfo{person}{Jason Weston}, {and} \bibinfo{person}{Oksana Yakhnenko}.} \bibinfo{year}{2013}\natexlab{}.
\newblock \showarticletitle{Translating Embeddings for Modeling Multi-relational Data}. In \bibinfo{booktitle}{\emph{Advances in Neural Information Processing Systems 26: 27th Annual Conference on Neural Information Processing Systems 2013. Proceedings of a meeting held December 5-8, 2013, Lake Tahoe, Nevada, United States}}. \bibinfo{pages}{2787--2795}.
\newblock
\urldef\tempurl%
\url{https://proceedings.neurips.cc/paper/2013/hash/1cecc7a77928ca8133fa24680a88d2f9-Abstract.html}
\showURL{%
\tempurl}


\bibitem[Broscheit et~al\mbox{.}(2020)]%
        {LibKGE}
\bibfield{author}{\bibinfo{person}{Samuel Broscheit}, \bibinfo{person}{Daniel Ruffinelli}, \bibinfo{person}{Adrian Kochsiek}, \bibinfo{person}{Patrick Betz}, {and} \bibinfo{person}{Rainer Gemulla}.} \bibinfo{year}{2020}\natexlab{}.
\newblock \showarticletitle{LibKGE - {A} knowledge graph embedding library for reproducible research}. In \bibinfo{booktitle}{\emph{Proceedings of the 2020 Conference on Empirical Methods in Natural Language Processing: System Demonstrations, {EMNLP} 2020 - Demos, Online, November 16-20, 2020}}. \bibinfo{pages}{165--174}.
\newblock
\href{https://doi.org/10.18653/V1/2020.EMNLP-DEMOS.22}{doi:\nolinkurl{10.18653/V1/2020.EMNLP-DEMOS.22}}


\bibitem[Cai et~al\mbox{.}(2023)]%
        {LightGCL}
\bibfield{author}{\bibinfo{person}{Xuheng Cai}, \bibinfo{person}{Chao Huang}, \bibinfo{person}{Lianghao Xia}, {and} \bibinfo{person}{Xubin Ren}.} \bibinfo{year}{2023}\natexlab{}.
\newblock \showarticletitle{LightGCL: Simple Yet Effective Graph Contrastive Learning for Recommendation}. In \bibinfo{booktitle}{\emph{The Eleventh International Conference on Learning Representations, {ICLR} 2023, Kigali, Rwanda, May 1-5, 2023}}.
\newblock
\urldef\tempurl%
\url{https://openreview.net/forum?id=FKXVK9dyMM}
\showURL{%
\tempurl}


\bibitem[Cai et~al\mbox{.}(2016)]%
        {GRETA}
\bibfield{author}{\bibinfo{person}{Xuyang Cai}, \bibinfo{person}{Jiangang Zhu}, \bibinfo{person}{Beijun Shen}, {and} \bibinfo{person}{Yuting Chen}.} \bibinfo{year}{2016}\natexlab{}.
\newblock \showarticletitle{{GRETA:} Graph-Based Tag Assignment for GitHub Repositories}. In \bibinfo{booktitle}{\emph{40th {IEEE} Annual Computer Software and Applications Conference, {COMPSAC} 2016, Atlanta, GA, USA, June 10-14, 2016}}. \bibinfo{pages}{63--72}.
\newblock
\href{https://doi.org/10.1109/COMPSAC.2016.124}{doi:\nolinkurl{10.1109/COMPSAC.2016.124}}


\bibitem[Cao et~al\mbox{.}(2021)]%
        {DEKR}
\bibfield{author}{\bibinfo{person}{Xianshuai Cao}, \bibinfo{person}{Yuliang Shi}, \bibinfo{person}{Han Yu}, \bibinfo{person}{Jihu Wang}, \bibinfo{person}{Xinjun Wang}, \bibinfo{person}{Zhongmin Yan}, {and} \bibinfo{person}{Zhiyong Chen}.} \bibinfo{year}{2021}\natexlab{}.
\newblock \showarticletitle{{DEKR:} Description Enhanced Knowledge Graph for Machine Learning Method Recommendation}. In \bibinfo{booktitle}{\emph{{SIGIR} '21: The 44th International {ACM} {SIGIR} Conference on Research and Development in Information Retrieval, Virtual Event, Canada, July 11-15, 2021}}. \bibinfo{pages}{203--212}.
\newblock
\href{https://doi.org/10.1145/3404835.3462900}{doi:\nolinkurl{10.1145/3404835.3462900}}


\bibitem[Cen et~al\mbox{.}(2023)]%
        {CogDL}
\bibfield{author}{\bibinfo{person}{Yukuo Cen}, \bibinfo{person}{Zhenyu Hou}, \bibinfo{person}{Yan Wang}, \bibinfo{person}{Qibin Chen}, \bibinfo{person}{Yizhen Luo}, \bibinfo{person}{Zhongming Yu}, \bibinfo{person}{Hengrui Zhang}, \bibinfo{person}{Xingcheng Yao}, \bibinfo{person}{Aohan Zeng}, \bibinfo{person}{Shiguang Guo}, \bibinfo{person}{Yuxiao Dong}, \bibinfo{person}{Yang Yang}, \bibinfo{person}{Peng Zhang}, \bibinfo{person}{Guohao Dai}, \bibinfo{person}{Yu Wang}, \bibinfo{person}{Chang Zhou}, \bibinfo{person}{Hongxia Yang}, {and} \bibinfo{person}{Jie Tang}.} \bibinfo{year}{2023}\natexlab{}.
\newblock \showarticletitle{CogDL: {A} Comprehensive Library for Graph Deep Learning}. In \bibinfo{booktitle}{\emph{Proceedings of the {ACM} Web Conference 2023, {WWW} 2023, Austin, TX, USA, 30 April 2023 - 4 May 2023}}. \bibinfo{pages}{747--758}.
\newblock
\href{https://doi.org/10.1145/3543507.3583472}{doi:\nolinkurl{10.1145/3543507.3583472}}


\bibitem[Chapman et~al\mbox{.}(2020)]%
        {DatasetSearchSurvey}
\bibfield{author}{\bibinfo{person}{Adriane Chapman}, \bibinfo{person}{Elena Simperl}, \bibinfo{person}{Laura Koesten}, \bibinfo{person}{George Konstantinidis}, \bibinfo{person}{Luis{-}Daniel Ib{\'{a}}{\~{n}}ez}, \bibinfo{person}{Emilia Kacprzak}, {and} \bibinfo{person}{Paul Groth}.} \bibinfo{year}{2020}\natexlab{}.
\newblock \showarticletitle{Dataset search: a survey}.
\newblock \bibinfo{journal}{\emph{{VLDB} J.}} \bibinfo{volume}{29}, \bibinfo{number}{1} (\bibinfo{year}{2020}), \bibinfo{pages}{251--272}.
\newblock
\href{https://doi.org/10.1007/S00778-019-00564-X}{doi:\nolinkurl{10.1007/S00778-019-00564-X}}


\bibitem[Chen et~al\mbox{.}(2020)]%
        {GCNII}
\bibfield{author}{\bibinfo{person}{Ming Chen}, \bibinfo{person}{Zhewei Wei}, \bibinfo{person}{Zengfeng Huang}, \bibinfo{person}{Bolin Ding}, {and} \bibinfo{person}{Yaliang Li}.} \bibinfo{year}{2020}\natexlab{}.
\newblock \showarticletitle{Simple and Deep Graph Convolutional Networks}. In \bibinfo{booktitle}{\emph{Proceedings of the 37th International Conference on Machine Learning, {ICML} 2020, 13-18 July 2020, Virtual Event}} \emph{(\bibinfo{series}{Proceedings of Machine Learning Research}, Vol.~\bibinfo{volume}{119})}. \bibinfo{pages}{1725--1735}.
\newblock
\urldef\tempurl%
\url{http://proceedings.mlr.press/v119/chen20v.html}
\showURL{%
\tempurl}


\bibitem[Chen et~al\mbox{.}(2021)]%
        {HittER}
\bibfield{author}{\bibinfo{person}{Sanxing Chen}, \bibinfo{person}{Xiaodong Liu}, \bibinfo{person}{Jianfeng Gao}, \bibinfo{person}{Jian Jiao}, \bibinfo{person}{Ruofei Zhang}, {and} \bibinfo{person}{Yangfeng Ji}.} \bibinfo{year}{2021}\natexlab{}.
\newblock \showarticletitle{HittER: Hierarchical Transformers for Knowledge Graph Embeddings}. In \bibinfo{booktitle}{\emph{Proceedings of the 2021 Conference on Empirical Methods in Natural Language Processing, {EMNLP} 2021, Virtual Event / Punta Cana, Dominican Republic, 7-11 November, 2021}}. \bibinfo{pages}{10395--10407}.
\newblock
\href{https://doi.org/10.18653/V1/2021.EMNLP-MAIN.812}{doi:\nolinkurl{10.18653/V1/2021.EMNLP-MAIN.812}}


\bibitem[Cui et~al\mbox{.}(2024)]%
        {KG-ICL}
\bibfield{author}{\bibinfo{person}{Yuanning Cui}, \bibinfo{person}{Zequn Sun}, {and} \bibinfo{person}{Wei Hu}.} \bibinfo{year}{2024}\natexlab{}.
\newblock \showarticletitle{A Prompt-Based Knowledge Graph Foundation Model for Universal In-Context Reasoning}. In \bibinfo{booktitle}{\emph{Advances in Neural Information Processing Systems 38: Annual Conference on Neural Information Processing Systems 2024, NeurIPS 2024, Vancouver, BC, Canada, December 10 - 15, 2024}}.
\newblock
\urldef\tempurl%
\url{http://papers.nips.cc/paper\_files/paper/2024/hash/0d70af566e69f1dfb687791ecf955e28-Abstract-Conference.html}
\showURL{%
\tempurl}


\bibitem[Dettmers et~al\mbox{.}(2018a)]%
        {ConvE}
\bibfield{author}{\bibinfo{person}{Tim Dettmers}, \bibinfo{person}{Pasquale Minervini}, \bibinfo{person}{Pontus Stenetorp}, {and} \bibinfo{person}{Sebastian Riedel}.} \bibinfo{year}{2018}\natexlab{a}.
\newblock \showarticletitle{Convolutional 2D Knowledge Graph Embeddings}. In \bibinfo{booktitle}{\emph{Proceedings of the Thirty-Second {AAAI} Conference on Artificial Intelligence, (AAAI-18), the 30th innovative Applications of Artificial Intelligence (IAAI-18), and the 8th {AAAI} Symposium on Educational Advances in Artificial Intelligence (EAAI-18), New Orleans, Louisiana, USA, February 2-7, 2018}}. \bibinfo{pages}{1811--1818}.
\newblock
\href{https://doi.org/10.1609/AAAI.V32I1.11573}{doi:\nolinkurl{10.1609/AAAI.V32I1.11573}}


\bibitem[Dettmers et~al\mbox{.}(2018b)]%
        {WN18RR}
\bibfield{author}{\bibinfo{person}{Tim Dettmers}, \bibinfo{person}{Pasquale Minervini}, \bibinfo{person}{Pontus Stenetorp}, {and} \bibinfo{person}{Sebastian Riedel}.} \bibinfo{year}{2018}\natexlab{b}.
\newblock \showarticletitle{Convolutional 2D Knowledge Graph Embeddings}. In \bibinfo{booktitle}{\emph{Proceedings of the Thirty-Second {AAAI} Conference on Artificial Intelligence, (AAAI-18), the 30th innovative Applications of Artificial Intelligence (IAAI-18), and the 8th {AAAI} Symposium on Educational Advances in Artificial Intelligence (EAAI-18), New Orleans, Louisiana, USA, February 2-7, 2018}}. \bibinfo{publisher}{{AAAI} Press}, \bibinfo{pages}{1811--1818}.
\newblock
\href{https://doi.org/10.1609/AAAI.V32I1.11573}{doi:\nolinkurl{10.1609/AAAI.V32I1.11573}}


\bibitem[Feng et~al\mbox{.}(2020)]%
        {GRAND}
\bibfield{author}{\bibinfo{person}{Wenzheng Feng}, \bibinfo{person}{Jie Zhang}, \bibinfo{person}{Yuxiao Dong}, \bibinfo{person}{Yu Han}, \bibinfo{person}{Huanbo Luan}, \bibinfo{person}{Qian Xu}, \bibinfo{person}{Qiang Yang}, \bibinfo{person}{Evgeny Kharlamov}, {and} \bibinfo{person}{Jie Tang}.} \bibinfo{year}{2020}\natexlab{}.
\newblock \showarticletitle{Graph Random Neural Networks for Semi-Supervised Learning on Graphs}. In \bibinfo{booktitle}{\emph{Advances in Neural Information Processing Systems 33: Annual Conference on Neural Information Processing Systems 2020, NeurIPS 2020, December 6-12, 2020, virtual}}.
\newblock
\urldef\tempurl%
\url{https://proceedings.neurips.cc/paper/2020/hash/fb4c835feb0a65cc39739320d7a51c02-Abstract.html}
\showURL{%
\tempurl}


\bibitem[Galkin et~al\mbox{.}(2024)]%
        {ULTRA}
\bibfield{author}{\bibinfo{person}{Mikhail Galkin}, \bibinfo{person}{Xinyu Yuan}, \bibinfo{person}{Hesham Mostafa}, \bibinfo{person}{Jian Tang}, {and} \bibinfo{person}{Zhaocheng Zhu}.} \bibinfo{year}{2024}\natexlab{}.
\newblock \showarticletitle{Towards Foundation Models for Knowledge Graph Reasoning}. In \bibinfo{booktitle}{\emph{The Twelfth International Conference on Learning Representations, {ICLR} 2024, Vienna, Austria, May 7-11, 2024}}.
\newblock
\urldef\tempurl%
\url{https://openreview.net/forum?id=jVEoydFOl9}
\showURL{%
\tempurl}


\bibitem[Hamilton et~al\mbox{.}(2017)]%
        {GraphSAGE}
\bibfield{author}{\bibinfo{person}{William~L. Hamilton}, \bibinfo{person}{Zhitao Ying}, {and} \bibinfo{person}{Jure Leskovec}.} \bibinfo{year}{2017}\natexlab{}.
\newblock \showarticletitle{Inductive Representation Learning on Large Graphs}. In \bibinfo{booktitle}{\emph{Advances in Neural Information Processing Systems 30: Annual Conference on Neural Information Processing Systems 2017, December 4-9, 2017, Long Beach, CA, {USA}}}. \bibinfo{pages}{1024--1034}.
\newblock
\urldef\tempurl%
\url{https://proceedings.neurips.cc/paper/2017/hash/5dd9db5e033da9c6fb5ba83c7a7ebea9-Abstract.html}
\showURL{%
\tempurl}


\bibitem[He et~al\mbox{.}(2020)]%
        {LightGCN}
\bibfield{author}{\bibinfo{person}{Xiangnan He}, \bibinfo{person}{Kuan Deng}, \bibinfo{person}{Xiang Wang}, \bibinfo{person}{Yan Li}, \bibinfo{person}{Yong{-}Dong Zhang}, {and} \bibinfo{person}{Meng Wang}.} \bibinfo{year}{2020}\natexlab{}.
\newblock \showarticletitle{LightGCN: Simplifying and Powering Graph Convolution Network for Recommendation}. In \bibinfo{booktitle}{\emph{Proceedings of the 43rd International {ACM} {SIGIR} conference on research and development in Information Retrieval, {SIGIR} 2020, Virtual Event, China, July 25-30, 2020}}. \bibinfo{pages}{639--648}.
\newblock
\href{https://doi.org/10.1145/3397271.3401063}{doi:\nolinkurl{10.1145/3397271.3401063}}


\bibitem[Hu et~al\mbox{.}(2020)]%
        {obgn-arxiv}
\bibfield{author}{\bibinfo{person}{Weihua Hu}, \bibinfo{person}{Matthias Fey}, \bibinfo{person}{Marinka Zitnik}, \bibinfo{person}{Yuxiao Dong}, \bibinfo{person}{Hongyu Ren}, \bibinfo{person}{Bowen Liu}, \bibinfo{person}{Michele Catasta}, {and} \bibinfo{person}{Jure Leskovec}.} \bibinfo{year}{2020}\natexlab{}.
\newblock \showarticletitle{Open Graph Benchmark: Datasets for Machine Learning on Graphs}. In \bibinfo{booktitle}{\emph{Advances in Neural Information Processing Systems 33: Annual Conference on Neural Information Processing Systems 2020, NeurIPS 2020, December 6-12, 2020, virtual}}.
\newblock


\bibitem[Kipf and Welling(2017)]%
        {GCN}
\bibfield{author}{\bibinfo{person}{Thomas~N. Kipf} {and} \bibinfo{person}{Max Welling}.} \bibinfo{year}{2017}\natexlab{}.
\newblock \showarticletitle{Semi-Supervised Classification with Graph Convolutional Networks}. In \bibinfo{booktitle}{\emph{5th International Conference on Learning Representations, {ICLR} 2017, Toulon, France, April 24-26, 2017, Conference Track Proceedings}}.
\newblock
\urldef\tempurl%
\url{https://openreview.net/forum?id=SJU4ayYgl}
\showURL{%
\tempurl}


\bibitem[Klicpera et~al\mbox{.}(2019)]%
        {APPNP}
\bibfield{author}{\bibinfo{person}{Johannes Klicpera}, \bibinfo{person}{Aleksandar Bojchevski}, {and} \bibinfo{person}{Stephan G{\"{u}}nnemann}.} \bibinfo{year}{2019}\natexlab{}.
\newblock \showarticletitle{Predict then Propagate: Graph Neural Networks meet Personalized PageRank}. In \bibinfo{booktitle}{\emph{7th International Conference on Learning Representations, {ICLR} 2019, New Orleans, LA, USA, May 6-9, 2019}}.
\newblock
\urldef\tempurl%
\url{https://openreview.net/forum?id=H1gL-2A9Ym}
\showURL{%
\tempurl}


\bibitem[Lhoest et~al\mbox{.}(2021)]%
        {datasets}
\bibfield{author}{\bibinfo{person}{Quentin Lhoest}, \bibinfo{person}{Albert~Villanova del Moral}, \bibinfo{person}{Yacine Jernite}, \bibinfo{person}{Abhishek Thakur}, \bibinfo{person}{Patrick von Platen}, \bibinfo{person}{Suraj Patil}, \bibinfo{person}{Julien Chaumond}, \bibinfo{person}{Mariama Drame}, \bibinfo{person}{Julien Plu}, \bibinfo{person}{Lewis Tunstall}, \bibinfo{person}{Joe Davison}, \bibinfo{person}{Mario Sasko}, \bibinfo{person}{Gunjan Chhablani}, \bibinfo{person}{Bhavitvya Malik}, \bibinfo{person}{Simon Brandeis}, \bibinfo{person}{Teven~Le Scao}, \bibinfo{person}{Victor Sanh}, \bibinfo{person}{Canwen Xu}, \bibinfo{person}{Nicolas Patry}, \bibinfo{person}{Angelina McMillan{-}Major}, \bibinfo{person}{Philipp Schmid}, \bibinfo{person}{Sylvain Gugger}, \bibinfo{person}{Cl{\'{e}}ment Delangue}, \bibinfo{person}{Th{\'{e}}o Matussi{\`{e}}re}, \bibinfo{person}{Lysandre Debut}, \bibinfo{person}{Stas Bekman}, \bibinfo{person}{Pierric Cistac}, \bibinfo{person}{Thibault Goehringer}, \bibinfo{person}{Victor
  Mustar}, \bibinfo{person}{Fran{\c{c}}ois Lagunas}, \bibinfo{person}{Alexander~M. Rush}, {and} \bibinfo{person}{Thomas Wolf}.} \bibinfo{year}{2021}\natexlab{}.
\newblock \showarticletitle{Datasets: {A} Community Library for Natural Language Processing}. In \bibinfo{booktitle}{\emph{Proceedings of the 2021 Conference on Empirical Methods in Natural Language Processing: System Demonstrations, {EMNLP} 2021, Online and Punta Cana, Dominican Republic, 7-11 November, 2021}}. \bibinfo{pages}{175--184}.
\newblock
\href{https://doi.org/10.18653/V1/2021.EMNLP-DEMO.21}{doi:\nolinkurl{10.18653/V1/2021.EMNLP-DEMO.21}}


\bibitem[Li et~al\mbox{.}(2021)]%
        {RevGNN}
\bibfield{author}{\bibinfo{person}{Guohao Li}, \bibinfo{person}{Matthias M{\"{u}}ller}, \bibinfo{person}{Bernard Ghanem}, {and} \bibinfo{person}{Vladlen Koltun}.} \bibinfo{year}{2021}\natexlab{}.
\newblock \showarticletitle{Training Graph Neural Networks with 1000 Layers}. In \bibinfo{booktitle}{\emph{Proceedings of the 38th International Conference on Machine Learning, {ICML} 2021, 18-24 July 2021, Virtual Event}} \emph{(\bibinfo{series}{Proceedings of Machine Learning Research}, Vol.~\bibinfo{volume}{139})}. \bibinfo{pages}{6437--6449}.
\newblock
\urldef\tempurl%
\url{http://proceedings.mlr.press/v139/li21o.html}
\showURL{%
\tempurl}


\bibitem[Liu et~al\mbox{.}(2023)]%
        {tse23}
\bibfield{author}{\bibinfo{person}{Mingwei Liu}, \bibinfo{person}{Chengyuan Zhao}, \bibinfo{person}{Xin Peng}, \bibinfo{person}{Simin Yu}, \bibinfo{person}{Haofen Wang}, {and} \bibinfo{person}{Chaofeng Sha}.} \bibinfo{year}{2023}\natexlab{}.
\newblock \showarticletitle{Task-Oriented {ML/DL} Library Recommendation Based on a Knowledge Graph}.
\newblock \bibinfo{journal}{\emph{{IEEE} Trans. Software Eng.}} \bibinfo{volume}{49}, \bibinfo{number}{8} (\bibinfo{year}{2023}), \bibinfo{pages}{4081--4096}.
\newblock
\href{https://doi.org/10.1109/TSE.2023.3285280}{doi:\nolinkurl{10.1109/TSE.2023.3285280}}


\bibitem[Longpre et~al\mbox{.}(2024)]%
        {DatasetAudit}
\bibfield{author}{\bibinfo{person}{Shayne Longpre}, \bibinfo{person}{Robert Mahari}, \bibinfo{person}{Anthony Chen}, \bibinfo{person}{Naana Obeng{-}Marnu}, \bibinfo{person}{Damien Sileo}, \bibinfo{person}{William Brannon}, \bibinfo{person}{Niklas Muennighoff}, \bibinfo{person}{Nathan Khazam}, \bibinfo{person}{Jad Kabbara}, \bibinfo{person}{Kartik Perisetla}, \bibinfo{person}{Xinyi Wu}, \bibinfo{person}{Enrico Shippole}, \bibinfo{person}{Kurt~D. Bollacker}, \bibinfo{person}{Tongshuang Wu}, \bibinfo{person}{Luis Villa}, \bibinfo{person}{Sandy Pentland}, {and} \bibinfo{person}{Sara Hooker}.} \bibinfo{year}{2024}\natexlab{}.
\newblock \showarticletitle{A large-scale audit of dataset licensing and attribution in {AI}}.
\newblock \bibinfo{journal}{\emph{Nat. Mac. Intell.}} \bibinfo{volume}{6}, \bibinfo{number}{8} (\bibinfo{year}{2024}), \bibinfo{pages}{975--987}.
\newblock
\href{https://doi.org/10.1038/S42256-024-00878-8}{doi:\nolinkurl{10.1038/S42256-024-00878-8}}


\bibitem[Nguyen et~al\mbox{.}(2024)]%
        {ESEM24}
\bibfield{author}{\bibinfo{person}{Phuong~T. Nguyen}, \bibinfo{person}{Juri~Di Rocco}, \bibinfo{person}{Claudio~Di Sipio}, \bibinfo{person}{Mudita Shakya}, \bibinfo{person}{Davide~Di Ruscio}, {and} \bibinfo{person}{Massimiliano~Di Penta}.} \bibinfo{year}{2024}\natexlab{}.
\newblock \showarticletitle{Automatic Categorization of GitHub Actions with Transformers and Few-shot Learning}. In \bibinfo{booktitle}{\emph{Proceedings of the 18th {ACM/IEEE} International Symposium on Empirical Software Engineering and Measurement, {ESEM} 2024, Barcelona, Spain, October 24-25, 2024}}, \bibfield{editor}{\bibinfo{person}{Xavier Franch}, \bibinfo{person}{Maya Daneva}, \bibinfo{person}{Silverio Mart{\'{\i}}nez{-}Fern{\'{a}}ndez}, {and} \bibinfo{person}{Luigi Quaranta}} (Eds.). \bibinfo{publisher}{{ACM}}, \bibinfo{pages}{468--474}.
\newblock
\href{https://doi.org/10.1145/3674805.3690752}{doi:\nolinkurl{10.1145/3674805.3690752}}


\bibitem[Ni et~al\mbox{.}(2019)]%
        {Amazon}
\bibfield{author}{\bibinfo{person}{Jianmo Ni}, \bibinfo{person}{Jiacheng Li}, {and} \bibinfo{person}{Julian~J. McAuley}.} \bibinfo{year}{2019}\natexlab{}.
\newblock \showarticletitle{Justifying Recommendations using Distantly-Labeled Reviews and Fine-Grained Aspects}. In \bibinfo{booktitle}{\emph{Proceedings of the 2019 Conference on Empirical Methods in Natural Language Processing and the 9th International Joint Conference on Natural Language Processing, {EMNLP-IJCNLP} 2019, Hong Kong, China, November 3-7, 2019}}. \bibinfo{publisher}{Association for Computational Linguistics}, \bibinfo{pages}{188--197}.
\newblock
\href{https://doi.org/10.18653/V1/D19-1018}{doi:\nolinkurl{10.18653/V1/D19-1018}}


\bibitem[Nickel et~al\mbox{.}(2011)]%
        {RESCAL}
\bibfield{author}{\bibinfo{person}{Maximilian Nickel}, \bibinfo{person}{Volker Tresp}, {and} \bibinfo{person}{Hans{-}Peter Kriegel}.} \bibinfo{year}{2011}\natexlab{}.
\newblock \showarticletitle{A Three-Way Model for Collective Learning on Multi-Relational Data}. In \bibinfo{booktitle}{\emph{Proceedings of the 28th International Conference on Machine Learning, {ICML} 2011, Bellevue, Washington, USA, June 28 - July 2, 2011}}. \bibinfo{pages}{809--816}.
\newblock
\urldef\tempurl%
\url{https://icml.cc/2011/papers/438\_icmlpaper.pdf}
\showURL{%
\tempurl}


\bibitem[Paton et~al\mbox{.}(2023)]%
        {DatasetDiscoverySurvey}
\bibfield{author}{\bibinfo{person}{Norman~W. Paton}, \bibinfo{person}{Jiaoyan Chen}, {and} \bibinfo{person}{Zhenyu Wu}.} \bibinfo{year}{2023}\natexlab{}.
\newblock \showarticletitle{Dataset Discovery and Exploration: A Survey}.
\newblock \bibinfo{journal}{\emph{ACM Comput. Surv.}} \bibinfo{volume}{56}, \bibinfo{number}{4} (\bibinfo{year}{2023}), \bibinfo{numpages}{37}~pages.
\newblock
\showISSN{0360-0300}
\href{https://doi.org/10.1145/3626521}{doi:\nolinkurl{10.1145/3626521}}


\bibitem[Ren et~al\mbox{.}(2024a)]%
        {RLMRec}
\bibfield{author}{\bibinfo{person}{Xubin Ren}, \bibinfo{person}{Wei Wei}, \bibinfo{person}{Lianghao Xia}, \bibinfo{person}{Lixin Su}, \bibinfo{person}{Suqi Cheng}, \bibinfo{person}{Junfeng Wang}, \bibinfo{person}{Dawei Yin}, {and} \bibinfo{person}{Chao Huang}.} \bibinfo{year}{2024}\natexlab{a}.
\newblock \showarticletitle{Representation learning with large language models for recommendation}. In \bibinfo{booktitle}{\emph{Proceedings of the ACM on Web Conference 2024}}. \bibinfo{pages}{3464--3475}.
\newblock


\bibitem[Ren et~al\mbox{.}(2024b)]%
        {SSLRec}
\bibfield{author}{\bibinfo{person}{Xubin Ren}, \bibinfo{person}{Lianghao Xia}, \bibinfo{person}{Yuhao Yang}, \bibinfo{person}{Wei Wei}, \bibinfo{person}{Tianle Wang}, \bibinfo{person}{Xuheng Cai}, {and} \bibinfo{person}{Chao Huang}.} \bibinfo{year}{2024}\natexlab{b}.
\newblock \showarticletitle{SSLRec: {A} Self-Supervised Learning Framework for Recommendation}. In \bibinfo{booktitle}{\emph{Proceedings of the 17th {ACM} International Conference on Web Search and Data Mining, {WSDM} 2024, Merida, Mexico, March 4-8, 2024}}. \bibinfo{pages}{567--575}.
\newblock
\href{https://doi.org/10.1145/3616855.3635814}{doi:\nolinkurl{10.1145/3616855.3635814}}


\bibitem[Ren et~al\mbox{.}(2023)]%
        {DCCF}
\bibfield{author}{\bibinfo{person}{Xubin Ren}, \bibinfo{person}{Lianghao Xia}, \bibinfo{person}{Jiashu Zhao}, \bibinfo{person}{Dawei Yin}, {and} \bibinfo{person}{Chao Huang}.} \bibinfo{year}{2023}\natexlab{}.
\newblock \showarticletitle{Disentangled Contrastive Collaborative Filtering}. In \bibinfo{booktitle}{\emph{Proceedings of the 46th International {ACM} {SIGIR} Conference on Research and Development in Information Retrieval, {SIGIR} 2023, Taipei, Taiwan, July 23-27, 2023}}. \bibinfo{pages}{1137--1146}.
\newblock
\href{https://doi.org/10.1145/3539618.3591665}{doi:\nolinkurl{10.1145/3539618.3591665}}


\bibitem[Shao et~al\mbox{.}(2020)]%
        {paper2repo}
\bibfield{author}{\bibinfo{person}{Huajie Shao}, \bibinfo{person}{Dachun Sun}, \bibinfo{person}{Jiahao Wu}, \bibinfo{person}{Zecheng Zhang}, \bibinfo{person}{Aston Zhang}, \bibinfo{person}{Shuochao Yao}, \bibinfo{person}{Shengzhong Liu}, \bibinfo{person}{Tianshi Wang}, \bibinfo{person}{Chao Zhang}, {and} \bibinfo{person}{Tarek~F. Abdelzaher}.} \bibinfo{year}{2020}\natexlab{}.
\newblock \showarticletitle{paper2repo: GitHub Repository Recommendation for Academic Papers}. In \bibinfo{booktitle}{\emph{{WWW} '20: The Web Conference 2020, Taipei, Taiwan, April 20-24, 2020}}. \bibinfo{pages}{629--639}.
\newblock
\href{https://doi.org/10.1145/3366423.3380145}{doi:\nolinkurl{10.1145/3366423.3380145}}


\bibitem[Shen et~al\mbox{.}(2023)]%
        {HuggingGPT}
\bibfield{author}{\bibinfo{person}{Yongliang Shen}, \bibinfo{person}{Kaitao Song}, \bibinfo{person}{Xu Tan}, \bibinfo{person}{Dongsheng Li}, \bibinfo{person}{Weiming Lu}, {and} \bibinfo{person}{Yueting Zhuang}.} \bibinfo{year}{2023}\natexlab{}.
\newblock \showarticletitle{HuggingGPT: solving AI tasks with chatgpt and its friends in hugging face}. In \bibinfo{booktitle}{\emph{Proceedings of the 37th International Conference on Neural Information Processing Systems}}.
\newblock


\bibitem[Sipio et~al\mbox{.}(2024)]%
        {EASE24}
\bibfield{author}{\bibinfo{person}{Claudio~Di Sipio}, \bibinfo{person}{Riccardo Rubei}, \bibinfo{person}{Juri~Di Rocco}, \bibinfo{person}{Davide~Di Ruscio}, {and} \bibinfo{person}{Phuong~T. Nguyen}.} \bibinfo{year}{2024}\natexlab{}.
\newblock \showarticletitle{Automated categorization of pre-trained models in software engineering: {A} case study with a Hugging Face dataset}. In \bibinfo{booktitle}{\emph{Proceedings of the 28th International Conference on Evaluation and Assessment in Software Engineering, {EASE} 2024, Salerno, Italy, June 18-21, 2024}}. \bibinfo{publisher}{{ACM}}, \bibinfo{pages}{351--356}.
\newblock
\href{https://doi.org/10.1145/3661167.3661215}{doi:\nolinkurl{10.1145/3661167.3661215}}


\bibitem[Sun et~al\mbox{.}(2019)]%
        {RotatE}
\bibfield{author}{\bibinfo{person}{Zhiqing Sun}, \bibinfo{person}{Zhi{-}Hong Deng}, \bibinfo{person}{Jian{-}Yun Nie}, {and} \bibinfo{person}{Jian Tang}.} \bibinfo{year}{2019}\natexlab{}.
\newblock \showarticletitle{RotatE: Knowledge Graph Embedding by Relational Rotation in Complex Space}. In \bibinfo{booktitle}{\emph{7th International Conference on Learning Representations, {ICLR} 2019, New Orleans, LA, USA, May 6-9, 2019}}.
\newblock
\urldef\tempurl%
\url{https://openreview.net/forum?id=HkgEQnRqYQ}
\showURL{%
\tempurl}


\bibitem[Toutanova and Chen(2015)]%
        {FB15k-237}
\bibfield{author}{\bibinfo{person}{Kristina Toutanova} {and} \bibinfo{person}{Danqi Chen}.} \bibinfo{year}{2015}\natexlab{}.
\newblock \showarticletitle{Observed versus latent features for knowledge base and text inference}. In \bibinfo{booktitle}{\emph{Proceedings of the 3rd Workshop on Continuous Vector Space Models and their Compositionality, {CVSC} 2015, Beijing, China, July 26-31, 2015}}. \bibinfo{publisher}{Association for Computational Linguistics}, \bibinfo{pages}{57--66}.
\newblock
\href{https://doi.org/10.18653/V1/W15-4007}{doi:\nolinkurl{10.18653/V1/W15-4007}}


\bibitem[Trouillon et~al\mbox{.}(2016)]%
        {ComplEX}
\bibfield{author}{\bibinfo{person}{Th{\'{e}}o Trouillon}, \bibinfo{person}{Johannes Welbl}, \bibinfo{person}{Sebastian Riedel}, \bibinfo{person}{{\'{E}}ric Gaussier}, {and} \bibinfo{person}{Guillaume Bouchard}.} \bibinfo{year}{2016}\natexlab{}.
\newblock \showarticletitle{Complex Embeddings for Simple Link Prediction}. In \bibinfo{booktitle}{\emph{Proceedings of the 33nd International Conference on Machine Learning, {ICML} 2016, New York City, NY, USA, June 19-24, 2016}} \emph{(\bibinfo{series}{{JMLR} Workshop and Conference Proceedings}, Vol.~\bibinfo{volume}{48})}. \bibinfo{pages}{2071--2080}.
\newblock
\urldef\tempurl%
\url{http://proceedings.mlr.press/v48/trouillon16.html}
\showURL{%
\tempurl}


\bibitem[Velickovic et~al\mbox{.}(2018)]%
        {GAT}
\bibfield{author}{\bibinfo{person}{Petar Velickovic}, \bibinfo{person}{Guillem Cucurull}, \bibinfo{person}{Arantxa Casanova}, \bibinfo{person}{Adriana Romero}, \bibinfo{person}{Pietro Li{\`{o}}}, {and} \bibinfo{person}{Yoshua Bengio}.} \bibinfo{year}{2018}\natexlab{}.
\newblock \showarticletitle{Graph Attention Networks}. In \bibinfo{booktitle}{\emph{6th International Conference on Learning Representations, {ICLR} 2018, Vancouver, BC, Canada, April 30 - May 3, 2018, Conference Track Proceedings}}.
\newblock
\urldef\tempurl%
\url{https://openreview.net/forum?id=rJXMpikCZ}
\showURL{%
\tempurl}


\bibitem[Vrandecic and Kr{\"{o}}tzsch(2014)]%
        {Wikidata}
\bibfield{author}{\bibinfo{person}{Denny Vrandecic} {and} \bibinfo{person}{Markus Kr{\"{o}}tzsch}.} \bibinfo{year}{2014}\natexlab{}.
\newblock \showarticletitle{Wikidata: a free collaborative knowledgebase}.
\newblock \bibinfo{journal}{\emph{Commun. {ACM}}} \bibinfo{volume}{57}, \bibinfo{number}{10} (\bibinfo{year}{2014}), \bibinfo{pages}{78--85}.
\newblock
\href{https://doi.org/10.1145/2629489}{doi:\nolinkurl{10.1145/2629489}}


\bibitem[Wang et~al\mbox{.}(2023)]%
        {DSL}
\bibfield{author}{\bibinfo{person}{Tianle Wang}, \bibinfo{person}{Lianghao Xia}, {and} \bibinfo{person}{Chao Huang}.} \bibinfo{year}{2023}\natexlab{}.
\newblock \showarticletitle{Denoised Self-Augmented Learning for Social Recommendation}. In \bibinfo{booktitle}{\emph{Proceedings of the Thirty-Second International Joint Conference on Artificial Intelligence, {IJCAI} 2023, 19th-25th August 2023, Macao, SAR, China}}. \bibinfo{pages}{2324--2331}.
\newblock
\href{https://doi.org/10.24963/IJCAI.2023/258}{doi:\nolinkurl{10.24963/IJCAI.2023/258}}


\bibitem[Wang et~al\mbox{.}(2021)]%
        {KGIN}
\bibfield{author}{\bibinfo{person}{Xiang Wang}, \bibinfo{person}{Tinglin Huang}, \bibinfo{person}{Dingxian Wang}, \bibinfo{person}{Yancheng Yuan}, \bibinfo{person}{Zhenguang Liu}, \bibinfo{person}{Xiangnan He}, {and} \bibinfo{person}{Tat{-}Seng Chua}.} \bibinfo{year}{2021}\natexlab{}.
\newblock \showarticletitle{Learning Intents behind Interactions with Knowledge Graph for Recommendation}. In \bibinfo{booktitle}{\emph{{WWW} '21: The Web Conference 2021, Virtual Event / Ljubljana, Slovenia, April 19-23, 2021}}. \bibinfo{pages}{878--887}.
\newblock
\href{https://doi.org/10.1145/3442381.3450133}{doi:\nolinkurl{10.1145/3442381.3450133}}


\bibitem[Xia et~al\mbox{.}(2023)]%
        {AutoCF}
\bibfield{author}{\bibinfo{person}{Lianghao Xia}, \bibinfo{person}{Chao Huang}, \bibinfo{person}{Chunzhen Huang}, \bibinfo{person}{Kangyi Lin}, \bibinfo{person}{Tao Yu}, {and} \bibinfo{person}{Ben Kao}.} \bibinfo{year}{2023}\natexlab{}.
\newblock \showarticletitle{Automated Self-Supervised Learning for Recommendation}. In \bibinfo{booktitle}{\emph{Proceedings of the {ACM} Web Conference 2023, {WWW} 2023, Austin, TX, USA, 30 April 2023 - 4 May 2023}}. \bibinfo{pages}{992--1002}.
\newblock
\href{https://doi.org/10.1145/3543507.3583336}{doi:\nolinkurl{10.1145/3543507.3583336}}


\bibitem[Xia et~al\mbox{.}(2022)]%
        {HCCF}
\bibfield{author}{\bibinfo{person}{Lianghao Xia}, \bibinfo{person}{Chao Huang}, \bibinfo{person}{Yong Xu}, \bibinfo{person}{Jiashu Zhao}, \bibinfo{person}{Dawei Yin}, {and} \bibinfo{person}{Jimmy~X. Huang}.} \bibinfo{year}{2022}\natexlab{}.
\newblock \showarticletitle{Hypergraph Contrastive Collaborative Filtering}. In \bibinfo{booktitle}{\emph{{SIGIR} '22: The 45th International {ACM} {SIGIR} Conference on Research and Development in Information Retrieval, Madrid, Spain, July 11 - 15, 2022}}. \bibinfo{pages}{70--79}.
\newblock
\href{https://doi.org/10.1145/3477495.3532058}{doi:\nolinkurl{10.1145/3477495.3532058}}


\bibitem[Xu et~al\mbox{.}(2023)]%
        {RepoRecommendation}
\bibfield{author}{\bibinfo{person}{Yueshen Xu}, \bibinfo{person}{Yuhong Jiang}, \bibinfo{person}{Xinkui Zhao}, \bibinfo{person}{Ying Li}, {and} \bibinfo{person}{Rui Li}.} \bibinfo{year}{2023}\natexlab{}.
\newblock \showarticletitle{Personalized Repository Recommendation Service for Developers with Multi-modal Features Learning}. In \bibinfo{booktitle}{\emph{{IEEE} International Conference on Web Services, {ICWS} 2023, Chicago, IL, USA, July 2-8, 2023}}. \bibinfo{pages}{455--464}.
\newblock
\href{https://doi.org/10.1109/ICWS60048.2023.00064}{doi:\nolinkurl{10.1109/ICWS60048.2023.00064}}


\bibitem[Yang et~al\mbox{.}(2015)]%
        {DistMult}
\bibfield{author}{\bibinfo{person}{Bishan Yang}, \bibinfo{person}{Wen{-}tau Yih}, \bibinfo{person}{Xiaodong He}, \bibinfo{person}{Jianfeng Gao}, {and} \bibinfo{person}{Li Deng}.} \bibinfo{year}{2015}\natexlab{}.
\newblock \showarticletitle{Embedding Entities and Relations for Learning and Inference in Knowledge Bases}. In \bibinfo{booktitle}{\emph{3rd International Conference on Learning Representations, {ICLR} 2015, San Diego, CA, USA, May 7-9, 2015, Conference Track Proceedings}}.
\newblock
\urldef\tempurl%
\url{http://arxiv.org/abs/1412.6575}
\showURL{%
\tempurl}


\bibitem[Yang et~al\mbox{.}(2023)]%
        {KGRec}
\bibfield{author}{\bibinfo{person}{Yuhao Yang}, \bibinfo{person}{Chao Huang}, \bibinfo{person}{Lianghao Xia}, {and} \bibinfo{person}{Chunzhen Huang}.} \bibinfo{year}{2023}\natexlab{}.
\newblock \showarticletitle{Knowledge Graph Self-Supervised Rationalization for Recommendation}. In \bibinfo{booktitle}{\emph{Proceedings of the 29th {ACM} {SIGKDD} Conference on Knowledge Discovery and Data Mining, {KDD} 2023, Long Beach, CA, USA, August 6-10, 2023}}. \bibinfo{pages}{3046--3056}.
\newblock
\href{https://doi.org/10.1145/3580305.3599400}{doi:\nolinkurl{10.1145/3580305.3599400}}


\bibitem[Yang et~al\mbox{.}(2022)]%
        {KGCL}
\bibfield{author}{\bibinfo{person}{Yuhao Yang}, \bibinfo{person}{Chao Huang}, \bibinfo{person}{Lianghao Xia}, {and} \bibinfo{person}{Chenliang Li}.} \bibinfo{year}{2022}\natexlab{}.
\newblock \showarticletitle{Knowledge Graph Contrastive Learning for Recommendation}. In \bibinfo{booktitle}{\emph{{SIGIR} '22: The 45th International {ACM} {SIGIR} Conference on Research and Development in Information Retrieval, Madrid, Spain, July 11 - 15, 2022}}. \bibinfo{pages}{1434--1443}.
\newblock
\href{https://doi.org/10.1145/3477495.3532009}{doi:\nolinkurl{10.1145/3477495.3532009}}


\bibitem[Yu et~al\mbox{.}(2021)]%
        {MHCN}
\bibfield{author}{\bibinfo{person}{Junliang Yu}, \bibinfo{person}{Hongzhi Yin}, \bibinfo{person}{Jundong Li}, \bibinfo{person}{Qinyong Wang}, \bibinfo{person}{Nguyen Quoc~Viet Hung}, {and} \bibinfo{person}{Xiangliang Zhang}.} \bibinfo{year}{2021}\natexlab{}.
\newblock \showarticletitle{Self-Supervised Multi-Channel Hypergraph Convolutional Network for Social Recommendation}. In \bibinfo{booktitle}{\emph{{WWW} '21: The Web Conference 2021, Virtual Event / Ljubljana, Slovenia, April 19-23, 2021}}. \bibinfo{pages}{413--424}.
\newblock
\href{https://doi.org/10.1145/3442381.3449844}{doi:\nolinkurl{10.1145/3442381.3449844}}


\bibitem[Yu et~al\mbox{.}(2022)]%
        {SimGCL}
\bibfield{author}{\bibinfo{person}{Junliang Yu}, \bibinfo{person}{Hongzhi Yin}, \bibinfo{person}{Xin Xia}, \bibinfo{person}{Tong Chen}, \bibinfo{person}{Lizhen Cui}, {and} \bibinfo{person}{Quoc Viet~Hung Nguyen}.} \bibinfo{year}{2022}\natexlab{}.
\newblock \showarticletitle{Are Graph Augmentations Necessary?: Simple Graph Contrastive Learning for Recommendation}. In \bibinfo{booktitle}{\emph{{SIGIR} '22: The 45th International {ACM} {SIGIR} Conference on Research and Development in Information Retrieval, Madrid, Spain, July 11 - 15, 2022}}. \bibinfo{pages}{1294--1303}.
\newblock
\href{https://doi.org/10.1145/3477495.3531937}{doi:\nolinkurl{10.1145/3477495.3531937}}


\bibitem[Zhang et~al\mbox{.}(2019)]%
        {OAG}
\bibfield{author}{\bibinfo{person}{Fanjin Zhang}, \bibinfo{person}{Xiao Liu}, \bibinfo{person}{Jie Tang}, \bibinfo{person}{Yuxiao Dong}, \bibinfo{person}{Peiran Yao}, \bibinfo{person}{Jie Zhang}, \bibinfo{person}{Xiaotao Gu}, \bibinfo{person}{Yan Wang}, \bibinfo{person}{Bin Shao}, \bibinfo{person}{Rui Li}, {and} \bibinfo{person}{Kuansan Wang}.} \bibinfo{year}{2019}\natexlab{}.
\newblock \showarticletitle{{OAG:} Toward Linking Large-scale Heterogeneous Entity Graphs}. In \bibinfo{booktitle}{\emph{Proceedings of the 25th {ACM} {SIGKDD} International Conference on Knowledge Discovery {\&} Data Mining, {KDD} 2019, Anchorage, AK, USA, August 4-8, 2019}}. \bibinfo{publisher}{{ACM}}, \bibinfo{pages}{2585--2595}.
\newblock
\href{https://doi.org/10.1145/3292500.3330785}{doi:\nolinkurl{10.1145/3292500.3330785}}


\bibitem[Zhang et~al\mbox{.}(2024)]%
        {OAGBench}
\bibfield{author}{\bibinfo{person}{Fanjin Zhang}, \bibinfo{person}{Shijie Shi}, \bibinfo{person}{Yifan Zhu}, \bibinfo{person}{Bo Chen}, \bibinfo{person}{Yukuo Cen}, \bibinfo{person}{Jifan Yu}, \bibinfo{person}{Yelin Chen}, \bibinfo{person}{Lulu Wang}, \bibinfo{person}{Qingfei Zhao}, \bibinfo{person}{Yuqing Cheng}, \bibinfo{person}{Tianyi Han}, \bibinfo{person}{Yuwei An}, \bibinfo{person}{Dan Zhang}, \bibinfo{person}{Weng~Lam Tam}, \bibinfo{person}{Kun Cao}, \bibinfo{person}{Yunhe Pang}, \bibinfo{person}{Xinyu Guan}, \bibinfo{person}{Huihui Yuan}, \bibinfo{person}{Jian Song}, \bibinfo{person}{Xiaoyan Li}, \bibinfo{person}{Yuxiao Dong}, {and} \bibinfo{person}{Jie Tang}.} \bibinfo{year}{2024}\natexlab{}.
\newblock \showarticletitle{OAG-Bench: {A} Human-Curated Benchmark for Academic Graph Mining}. In \bibinfo{booktitle}{\emph{Proceedings of the 30th {ACM} {SIGKDD} Conference on Knowledge Discovery and Data Mining, {KDD} 2024, Barcelona, Spain, August 25-29, 2024}}. \bibinfo{publisher}{{ACM}}, \bibinfo{pages}{6214--6225}.
\newblock
\href{https://doi.org/10.1145/3637528.3672354}{doi:\nolinkurl{10.1145/3637528.3672354}}


\end{thebibliography}

\end{document}